\documentclass[sigconf]{acmart}
\usepackage{balance} 
\usepackage{xspace}
\usepackage{color}
\usepackage{caption}
\usepackage{subcaption}
\usepackage{tikz}
\usepackage{pgf}
\usepackage{pgfplots}
\pgfplotsset{compat=newest}
\usepgfplotslibrary{units,external,groupplots}
\usetikzlibrary{positioning}
\usepackage{paralist}
\usepackage{pgfplotstable}
\usepackage{amsmath,amssymb}
\usepackage{wrapfig}
\usepackage{siunitx}
\usepackage[utf8]{inputenc}
\DeclareSIUnit\dBm{dBm}
\usetikzlibrary{patterns,angles,quotes,positioning,calc,shapes}
\usepackage{xfrac}

\graphicspath{{./img/}}

\usepackage{ifthen}
\newboolean{vspaceoptimization}
\setboolean{vspaceoptimization}{false}
\ifthenelse{\boolean{vspaceoptimization}}
{
\newcommand{\vsquish}[1]{\vspace{#1}}
}
{
\newcommand{\vsquish}[1]{}
}

\newboolean{authnotes}

\setboolean{authnotes}{false}

\ifthenelse{\boolean{authnotes}}
{

\newcommand{\fm}[1]{\footnote{{\bf\color{blue} FM: #1}}}
\newcommand{\mz}[1]{\footnote{{\bf\color{blue!50!orange} MZ: #1}}}
\newcommand{\db}[1]{\footnote{{\bf\color{red} DB: #1}}}
\newcommand{\st}[1]{\footnote{{\bf\color{red!70!yellow} ST: #1}}}
\newcommand{\rj}[1]{\footnote{{\bf\color{green!50!black} RJ: #1}}}
}
{
\newcommand{\fm}[1]{}
\newcommand{\db}[1]{}
\newcommand{\st}[1]{}
\newcommand{\mz}[1]{}
\newcommand{\rj}[1]{}
}

\newcommand\figref[1]{Figure~\ref{#1}}

\newcommand\secref[1]{Section~\ref{#1}}
\newcommand\appref[1]{Appendix~\ref{#1}}

\newcommand{\etal}{et~al.\xspace}
\newcommand{\eg}{\emph{e.g.},\xspace}
\newcommand{\ie}{\emph{i.e.},\xspace}
\newcommand{\etc}{etc.\xspace}
\newcommand{\capt}[1]{\mdseries{\emph{#1}}}

\newcommand{\fakepar}[1]{\vspace{1mm}\noindent\textbf{#1.}}

\newcommand{\cps}{CPS\xspace}

\newcommand{\ttw}{TTW\xspace}
\newcommand{\iid}{i.i.d.\xspace}

\newcommand{\ap}{AP\xspace}
\newcommand{\apnode}[1]{AP\textsubscript{#1}}
\newcommand{\cp}{CP\xspace}
\newcommand{\cpnode}[1]{CP\textsubscript{#1}}
\newcommand{\dpp}{DPP\xspace}
\newcommand{\bolt}{Bolt\xspace}

\newcommand{\sync}{SYNC\xspace}

\newcommand{\eREF}{\ensuremath{e_{\mathit{ref}}}\xspace}
\newcommand{\eREFmax}{\ensuremath{\hat{e}_{\mathit{ref}}}\xspace}

\newcommand{\eSYNCmax}{\ensuremath{\hat{e}_{\mathit{SYNC}}}\xspace}

\newcommand{\eTASKmax}{\ensuremath{\hat{e}_{\mathit{task}}}\xspace}

\newcommand{\rhoAP}{\ensuremath{\rho_{\mathit{\ap}}}\xspace}
\newcommand{\rhoAPmax}{\ensuremath{\hat{\rho}_{\mathit{\ap}}}\xspace}
\newcommand{\rhoCP}{\ensuremath{\rho_{\mathit{\cp}}}\xspace}
\newcommand{\rhoCPmax}{\ensuremath{\hat{\rho}_{\mathit{\cp}}}\xspace}

\newcommand{\trefEstim}{\ensuremath{\hat{t}_{\mathit{ref}}}\xspace}

\newcommand{\fAP}{\ensuremath{f_{\mathit{\ap}}}\xspace}

\newcommand{\Tjitter}{\ensuremath{\widetilde{T}_{\mathit{end}}}\xspace}
\newcommand{\Jitter}{\ensuremath{J}\xspace}

\newcommand{\Tupdate}{\ensuremath{T_U}\xspace}
\newcommand{\Tdelay}{\ensuremath{T_D}\xspace}

\newcommand{\meter}{\ensuremath{\,\text{m}}\xspace}
\newcommand{\s}{\ensuremath{\,\text{s}}\xspace}
\newcommand{\ms}{\ensuremath{\,\text{ms}}\xspace}

\newcommand{\MHz}{\ensuremath{\,\text{MHz}}\xspace}

\newcommand{\dBm}{\ensuremath{\,\text{dBm}}\xspace}
\newcommand{\ppm}{\ensuremath{\,\text{ppm}}\xspace}
\newcommand{\kbps}{\ensuremath{\,\text{kbit/s}}\xspace}

\newcommand{\percent}{\ensuremath{\,\text{\%}}\xspace}

\newcommand{\V}{\ensuremath{~\text{V}}\xspace}

\newtheorem{theo}{Theorem}
\newtheorem{lem}{Lemma}
\newtheorem{defi}{Definition}

\DeclareMathOperator*{\E}{\mathbb{E}}
\DeclareMathOperator*{\Var}{Var}

\author{Fabian Mager}
\authornote{Both authors contributed equally to this work.}
\affiliation{%
 \institution{TU Dresden}
}
\email{fabian.mager@tu-dresden.de}

\author{Dominik Baumann}
\authornotemark[1]
\affiliation{%
 \institution{MPI for Intelligent Systems}
}
\email{dominik.baumann@tuebingen.mpg.de}

\author{Romain Jacob}
\affiliation{%
 \institution{ETH Zurich}
}
\email{romain.jacob@tik.ee.ethz.ch}

\author{Lothar Thiele}
\affiliation{%
 \institution{ETH Zurich}
}
\email{thiele@ethz.ch}

\author{Sebastian Trimpe}
\affiliation{%
 \institution{MPI for Intelligent Systems}
}
\email{trimpe@is.mpg.de}

\author{Marco Zimmerling}
\affiliation{%
 \institution{TU Dresden}
}
\email{marco.zimmerling@tu-dresden.de}

\renewcommand{\shortauthors}{F.~Mager et al.}

\begin{document}

\copyrightyear{2019}
\acmYear{2019}
\setcopyright{acmlicensed}
\acmConference[ICCPS '19]{10th ACM/IEEE International Conference on Cyber-Physical Systems (with CPS-IoT Week 2019)}{April 16--18, 2019}{Montreal, QC, Canada}
\acmBooktitle{10th ACM/IEEE International Conference on Cyber-Physical Systems (with CPS-IoT Week 2019) (ICCPS '19), April 16--18, 2019, Montreal, QC, Canada}
\acmPrice{15.00}
\acmDOI{10.1145/3302509.3311046}
\acmISBN{978-1-4503-6285-6/19/04}

\setdefaultleftmargin{2em}{}{}{}{}{}

\title[Feedback Control Goes Wireless]{Feedback Control Goes Wireless: Guaranteed Stability over Low-power Multi-hop Networks}


\begin{abstract}
\vspace{-0.5mm}
Closing feedback loops fast and over long distances is key to emerging applications; for example, robot motion control and swarm coordination require update intervals of tens of milliseconds.
Low-power wireless technology is preferred for its low cost, small form factor, and flexibility, especially if the devices support multi-hop communication.
So far, however, feedback control over wireless multi-hop networks has only been shown for update intervals on the order of seconds.
This paper presents a wireless embedded system that tames imperfections impairing control performance (\eg jitter and message loss), and a control design that exploits the essential properties of this system to provably guarantee closed-loop stability for physical processes with linear time-invariant dynamics.
Using experiments on a cyber-physical testbed with 20 wireless nodes and multiple cart-pole systems, we are the first to demonstrate and evaluate feedback control and coordination over wireless multi-hop networks for update intervals of \mbox{20 to 50 milliseconds.}%
\vspace{-0.5mm}
\end{abstract}

\ccsdesc[500]{Computer systems organization~Sensors and actuators}
\ccsdesc[300]{Computer systems organization~Embedded systems}
\ccsdesc[300]{Computer systems organization~Real-time system architecture}
\ccsdesc[300]{Computer systems organization~Dependable and fault-tolerant systems and networks}
\ccsdesc[500]{Networks~Cyber-physical networks}
\ccsdesc[300]{Networks~Network protocol design}

\keywords{Wireless control, Closed-loop stability, Multi-agent systems, Multi-hop networks, Synchronous transmissions, Cyber-physical systems}

\maketitle


\section{Introduction}
\label{sec:intro}

Cyber-physical systems~(\cps) use embedded computers and networks to monitor and control physical systems~\cite{Derler2012}.
While monitoring using \emph{sensors} allows, for example, to better understand certain characteristics of environmental processes~\cite{Corke2010}, it is control and coordination through \emph{actuators} what nurtures the \cps vision exemplified by robotic materials~\cite{Correll2017}, smart transportation~\cite{Besselink2016}, and multi-robot swarms for disaster response and manufacturing~\cite{Hayat2016}.

A key hurdle to realizing this vision is how to close the \emph{feedback loops} between sensors and actuators as these may be numerous, mobile, distributed over large spaces, and attached to devices with size, weight, and cost constraints.
Wireless multi-hop communication among low-power, possibly battery-powered devices\footnote{While actuators may need wall power, low-power operation is crucial for sensors and controllers, which may run on batteries and harvest energy from the environment~\cite{Akerberg2011}.} provides the cost efficiency and flexibility to overcome this hurdle~\cite{Lu2016,Watteyne2016} if two requirements are met.
First, fast feedback is needed to keep up with the dynamics of physical systems~\cite{Astrom1996}; for example, robot motion control and drone swarm coordination require update intervals of tens of milliseconds~\cite{Preiss2017,Abbenseth2017}.
Second, as feedback control modifies the dynamics of physical systems~\cite{Astrom2008}, guaranteeing \emph{closed-loop stability} under imperfect wireless communication is a major concern. 

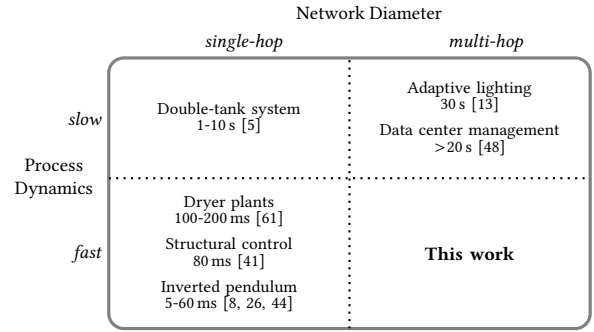
\begin{figure}[!tb]
\centering
\tikzsetnextfilename{cps_solutions}
\tikzexternalexportnextfalse
\begin{tikzpicture}[scale=0.8, every node/.style={scale=0.8}]]
\begin{scope}
\path[clip, preaction={draw, very thick, gray}] [rounded corners=2mm] (-4,-2.5) rectangle (4,2);
\draw[thick,dotted] (-4,0) -- (4,0);
\draw[thick,dotted] (0,-2.5) -- (0,2);
\end{scope}
\begin{scope}[text=black]
\node[align=center, anchor=east] at (-4.15,0) {Process\\Dynamics};
\node[align=right, anchor=east] at (-4,-1.25) {\textit{fast}};
\node[align=right, anchor=east] at (-4,1) {\textit{slow}};
%
\node[align=center, anchor=east] at (1.65,2.75) {Network Diameter};
\node[align=center, anchor=east] at (-1,2.25) {\textit{single-hop}};
\node[align=center, anchor=east] at (3,2.25) {\textit{multi-hop}};
\end{scope}
\begin{scope}[text width=4cm,text=black, align=center]
\node[font=\bfseries] (n1) at (2,-1.25) {This work};
\node[font=\small] (n2) at (-2,-1.25) {
	Dryer plants\\ \vspace{-0.08cm}
	\SIrange[range-units=single, range-phrase=-]{100}{200}{\milli\second} \cite{Ye2001}\\ \vspace{0.08cm}
	Structural control\\ \vspace{-0.08cm}
	\SI{80}{\milli\second} \cite{Lynch2008}\\ \vspace{0.08cm}
	Inverted pendulum\\ \vspace{-0.08cm}
	\SIrange[range-units=single, range-phrase=-]{5}{60}{\milli\second} \cite{Ploplys2004, Hernandez2011, Bauer2014}
};
\node[font=\small] (n3) at (-2,1) {
	Double-tank system\\ \vspace{-0.08cm}
	\SIrange[range-units=single, range-phrase=-]{1}{10}{\second} \cite{Araujo2014}
};
\node[font=\small] (n3) at (2,1) {
	Adaptive lighting\\ \vspace{-0.08cm}
	\SI{30}{\second} \cite{Ceriotti2011}\\ \vspace{0.08cm}
	Data center management\\ \vspace{-0.08cm}
	\SI{>20}{\second} \cite{Saifullah2014}
};
\end{scope}
\end{tikzpicture}
\vsquish{-2mm}
\caption{Design space of wireless \cps that have been validated on physical platforms and real wireless networks.}
\vsquish{-4mm}
\label{fig:solutions}
\end{figure}

Hence, this paper investigates the following question: \emph{Is it possible to enable fast feedback control and coordination across \textbf{real-world} multi-hop low-power wireless networks with formal guarantees on closed-loop stability?}
Prior works on control over wireless that validate their design through experiments on physical platforms do not provide an affirmative answer.
As shown in \figref{fig:solutions} and detailed in \secref{sec:related}, solutions based on \emph{multi-hop} communication have only been demonstrated for physical systems with \emph{slow} dynamics (\ie update intervals of seconds) and do not provide stability guarantees.
Practical solutions with stability guarantees for \emph{fast} process dynamics (\ie update intervals of tens of milliseconds as typical of, \eg mechanical systems) exist, but these are only applicable to \emph{single-hop} networks and therefore lack the scalability and flexibility required by many future \cps applications~\cite{Hayat2016,Luvisotto2017}.

\fakepar{Contribution and road-map}
This paper presents the design, analysis, and practical validation of a wireless \cps that fills this gap.
\secref{sec:overview} highlights the main challenges and corresponding system design goals we must achieve when closing feedback loops fast over wireless multi-hop networks.
Our approach is based on a careful co-design of the wireless embedded components (in terms of hardware and software) and the closed-loop control system, as described in Sections~\ref{sec:embedded} and~\ref{sec:control}.
More concretely, we tame typical wireless network imperfections, such as message loss and jitter,
so that they can be tackled by well-known control techniques or safely neglected.
As a result, our solution is amenable to a formal end-to-end analysis of all \cps components (\ie wireless embedded, control, and physical systems), which we exploit to provably guarantee closed-loop stability for physical systems with \emph{linear time-invariant (LTI)} dynamics.
Moreover, unlike prior work, our solution supports control and coordination of multiple physical systems out of the box, which is a key asset in many \cps applications~\cite{Hayat2016,Preiss2017,Abbenseth2017}.

To evaluate our design in \secref{sec:eval}, we have developed a cyber-physical testbed that consists of 20 wireless embedded devices forming a three-hop network and multiple cart-pole systems whose dynamics match a range of real-world mechanical systems~\cite{Astrom2008,Trimpe2012}.
As such, our testbed addresses an important need in \cps research~\cite{Lu2016}.
Our experiments reveal the following key findings:
(\emph{i}) two inverted pendulums can be safely stabilized by two remote controllers across the three-hop wireless network;
(\emph{ii}) the movement of five cart-poles can be synchronized reliably over the network;
(\emph{iii}) increasing message loss rates and update intervals can be tolerated at reduced control performance; and
(\emph{iv}) the experiments confirm our analyses.

In summary, this paper contributes the following:
\begin{itemize}
 \item We are the first to demonstrate feedback control and coordination across real multi-hop low-power wireless networks at update intervals of 20 to 50 milliseconds.
 \item We formally prove that our end-to-end \cps design guarantees closed-loop stability for physical systems with LTI dynamics.
 \item Experiments on a novel cyber-physical testbed show that our solution can stabilize and synchronize multiple inverted pendulums despite significant message loss.
\end{itemize}

\section{Related Work}
\label{sec:related}

Feedback control over wireless communication networks has been extensively studied.
For instance, the control community has investigated control design and stability analysis for wireless (and wired) networks based on different system architectures, delay models, and message loss processes (see, \eg~\cite{Luck1990,Sinopoli2004,Zhang2001,Xiong2007,Walsh2002,Gatsis2014,Quevedo2012,Alur2011,Smarra2012}); recent surveys provide an overview of this body of fundamental research~\cite{Hespanha2007,Zhang2013}.
However, the majority of those works focuses on theoretical analyses or validates new wireless \cps designs (\eg based on WirelessHART~\cite{Li2016,Ma2018}) only in simulation, thereby ignoring many fundamental challenges that may complicate or prevent a real implementation~\cite{Lu2016}.
One of the challenges, as detailed in \secref{sec:overview}, is that even slight variations in the quality of a wireless link can trigger drastic changes in the routing topology~\cite{Ceriotti2011}---and this can happen several times per minute~\cite{Gnawali2009}.
Hence, to establish trust in feedback control over wireless, a real-world validation against these \emph{dynamics} on a realistic \cps testbed is absolutely essential~\cite{Lu2016}, as opposed to considering setups with a \emph{statically configured} routing topology and only a few nodes on a desk as, for example, in~\cite{Schindler2017}.
%
%
%
%

\figref{fig:solutions} classifies prior control-over-wireless solutions that have been validated using experiments on physical platforms and against the dynamics of real wireless networks along two dimensions: the diameter of the network (\emph{single-hop} or \emph{multi-hop}) and the dynamics of the physical system (\emph{slow} or \emph{fast}).  While not representing absolute categories,
we use slow to refer to update intervals on the order of seconds, which is typically insufficient for feedback control of, for example, mechanical systems.

In the \emph{single-hop/slow} category, Ara{\'{u}}jo \etal \cite{Araujo2014} investigate resource efficiency of aperiodic control with closed-loop stability in a single-hop wireless network of IEEE 802.15.4 devices.
Using a double-tank system as the physical process, update intervals of one to ten seconds are sufficient.


A number of works in the \emph{single-hop/fast} class stabilize an inverted pendulum via a controller that communicates with a sensor-actuator node at the cart.
The update interval is 60\ms or less, and the interplay of control and network performance, as well as closed-loop stability are investigated for different wireless technologies: Bluetooth~\cite{Eker2001}, IEEE 802.11~\cite{Ploplys2004}, and IEEE 802.15.4~\cite{Bauer2014,Hernandez2011}.
Belonging to the same class, Ye \etal use three IEEE 802.11 nodes to control two dryer plants at update intervals of 100--200\ms~\cite{Ye2001}, and Lynch \etal use four proprietary wireless nodes to demonstrate control of a three-story test structure at an update interval of 80\ms~\cite{Lynch2008}.

For \emph{multi-hop} networks, there are only solutions for \emph{slow} process dynamics and without stability analysis.
For example, Ceriotti \etal study adaptive lighting in road tunnels~\cite{Ceriotti2011}.
Owing to the length of the tunnels, multi-hop communication becomes unavoidable, yet the required update interval of 30 seconds allows for a reliable solution built out of mainstream sensor network technology.
Similarly, Saifullah \etal present a multi-hop solution for power management in data centers, using update intervals of 20 seconds or greater~\cite{Saifullah2014}.

In contrast to these works, as illustrated in \figref{fig:solutions}, we demonstrate \emph{fast} feedback control over wireless \emph{multi-hop} networks at update intervals of 20--50\ms, which is
significantly faster than existing multi-hop solutions.
Moreover, we provide a formal stability proof, and our solution seamlessly supports both control and coordination of multiple physical systems, which we validate through experiments on a real-world cyber-physical testbed.%

\section{Problem Setting and Approach}
\label{sec:overview}


\begin{figure}[!tb]
	\centering
	\includegraphics[width=\linewidth]{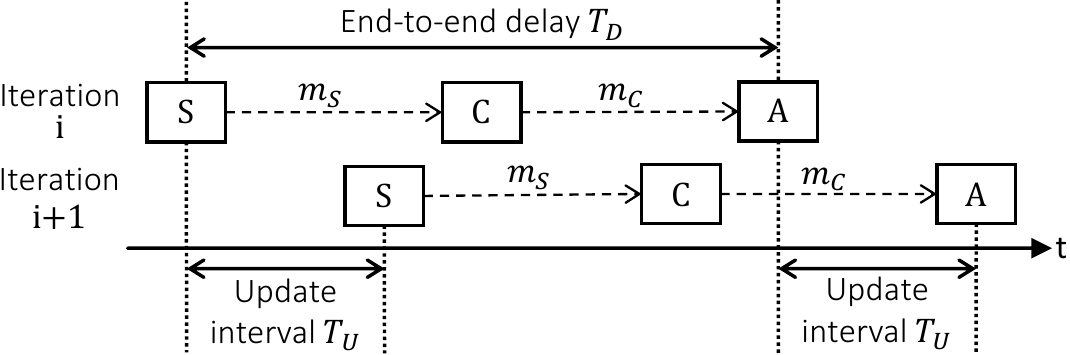}
	\vspace{-4mm}
	\caption{Application tasks and message transfers of a single feedback loop.
	\capt{In every iteration, the sensing task (S) takes a measurement of the physical system and sends it to the control task (C), which computes a control signal and sends it to the actuation task~(A).}}
	\vsquish{-6mm}
	\label{fig:overview}
\end{figure}

\vspace{-1mm}
\fakepar{Scenario}
We consider wireless \cps that consist of a set of embedded devices equipped with low-power wireless radios.
The devices execute different \emph{application tasks} (\ie sensing, control, and actuation) that exchange \emph{messages} over a wireless multi-hop network.
Each node may execute multiple application tasks, which may belong to different distributed feedback loops.
As an example, \figref{fig:overview}
shows the execution of application tasks and the exchange of messages for a single periodic feedback loop with one sensor and one actuator.
The \emph{update interval} \Tupdate is the time between consecutive sensing or actuation tasks.
The \emph{end-to-end delay} \Tdelay is the time between corresponding sensing and actuation tasks.

\fakepar{Challenges}
Fast feedback control over wireless multi-hop networks is an open problem due to the following challenges:
\begin{itemize}
 \item \emph{Lower end-to-end throughput.} Multi-hop networks have a lower end-to-end throughput than single-hop networks because of interference: the theoretical multi-hop upper bound is half the single-hop upper bound~\cite{Osterlind2008}. This limits the number of sensors and actuators that can be supported for a given maximum update interval.
 \item \emph{Significant delays and jitter.} Multi-hop networks also incur longer end-to-end delays, and the delays are subject to larger variations because of retransmissions or routing dynamics~\cite{Ceriotti2011}, introducing significant jitter. Delays and jitter can both destabilize a feedback system~\cite{Wittenmark1995,Walsh2002}.
 \item \emph{Constrained traffic patterns.} In a single-hop network, each node can communicate with every other node due to the broadcast property of the wireless medium.
 This is generally not the case in a multi-hop network. For example, WirelessHART only supports communication to and from a gateway that connects the wireless network to the control system. Feedback control under constrained traffic patterns is more challenging and may imply poor control performance or even infeasibility of closed-loop stability~\cite{Yang2005}.
 \item \emph{Correlated message loss.} Message loss is a common phenomenon in wireless networks, which complicates control design. Further, because there is often significant correlation among the message losses~\cite{Srinivasan2008}, a valid theoretical analysis to provide strong guarantees is hard, if not impossible.
 \item \emph{Message duplicates and out-of-order message delivery} are typical in wireless multi-hop protocols~\cite{Gnawali2009,Duquennoy2015} and may further hinder control design and stability analysis~\cite{Zhang2013}.
\end{itemize}

\vsquish{-1mm}
\fakepar{Approach}
We adopt the following co-design approach to enable fast feedback control over wireless multi-hop networks: \emph{Address the challenges on the wireless embedded system side to the extent possible, and then consider the resulting key properties in the control design.}
This entails the design of a wireless embedded system \mbox{that aims to:}
\begin{itemize}
 \item[\textbf{G1}] reduce and bound imperfections impairing control performance (\eg reduce \Tupdate and \Tdelay and bound their jitter);
 \item[\textbf{G2}] support arbitrary traffic patterns in multi-hop networks with real dynamics (\eg time-varying link qualities);
 \item[\textbf{G3}] operate efficiently in terms of limited resources, while accommodating the computational needs of the controller.
\end{itemize}
On the other hand, the control design aims to:
\begin{itemize}
 \item[\textbf{G4}] incorporate all essential properties of the wireless embedded system to guarantee closed-loop stability for the entire \cps for physical systems with LTI dynamics;
 \item[\textbf{G5}] enable an efficient implementation of the control logic on state-of-the-art low-power embedded devices;
 \item[\textbf{G6}] exploit the support for arbitrary traffic patterns for straightforward distributed control and multi-agent coordination.
\end{itemize}


\vsquish{-2mm}
\section{Wireless Embedded System Design}
\label{sec:embedded}

To achieve goals \textbf{G1}--\textbf{G3}, we design a wireless embedded system that consists of three key building blocks:
\begin{itemize}
 \item[1)] a \emph{low-power wireless protocol} that provides multi-hop many-to-all communication with minimal, bounded end-to-end delay and accurate network-wide time synchronization;
 \item[2)] a \emph{hardware platform} that enables a predictable and efficient execution of all application tasks and message transfers;
 \item[3)] a \emph{scheduling framework} to schedule all application tasks and message transfers so that given bounds on \Tupdate and \Tdelay are met at minimum energy costs for wireless communication.
\end{itemize}
We describe each building block below, followed by an analysis of the resulting properties that matter for the control design.%

\vsquish{-3mm}
\subsection{Low-power Wireless Protocol}
\label{sec:wireless_protocol}
\vsquish{-1mm}

To support arbitrary traffic patterns (\textbf{G2}), we require a multi-hop protocol capable of many-to-all communication.
Moreover, the protocol must be highly reliable and the time needed for many-to-all communication must be tightly bounded~(\textbf{G1}).
It has been shown that a solution based on Glossy floods~\cite{Ferrari2011} can meet these requirements with high efficiency (\textbf{G3}) in the face of wireless dynamics (\textbf{G2})~\cite{Zimmerling2017}.
Thus, similar to other recent proposals~\cite{Ferrari2012,Istomin2016}, we design a wireless protocol on top of Glossy, but aim at a new design point:
bounded end-to-end delays of at most a few tens of milliseconds \emph{for the many-to-all exchange of multiple messages} in a control cycle.

\begin{figure}[!tb]
	\centering
	\includegraphics[width=\linewidth]{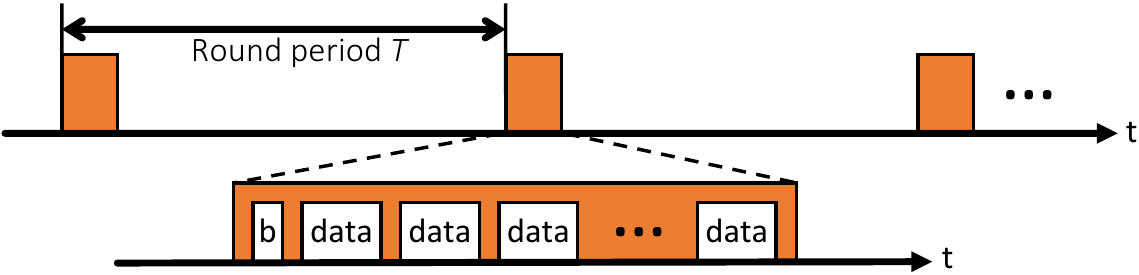}
	\vspace{-4mm}
	\caption{Time-triggered operation of multi-hop low-power wireless protocol. \capt{Communication occurs in rounds that are scheduled with a given round period $T$. Every beacon (b) and data slot in a round corresponds to an efficient, reliable one-to-all Glossy flood~\cite{Ferrari2011}.}}
	\vsquish{-4mm}
	\label{fig:protocol}
\end{figure}

\begin{figure*}[!tb]
	\centering
	\includegraphics[width=0.9\linewidth]{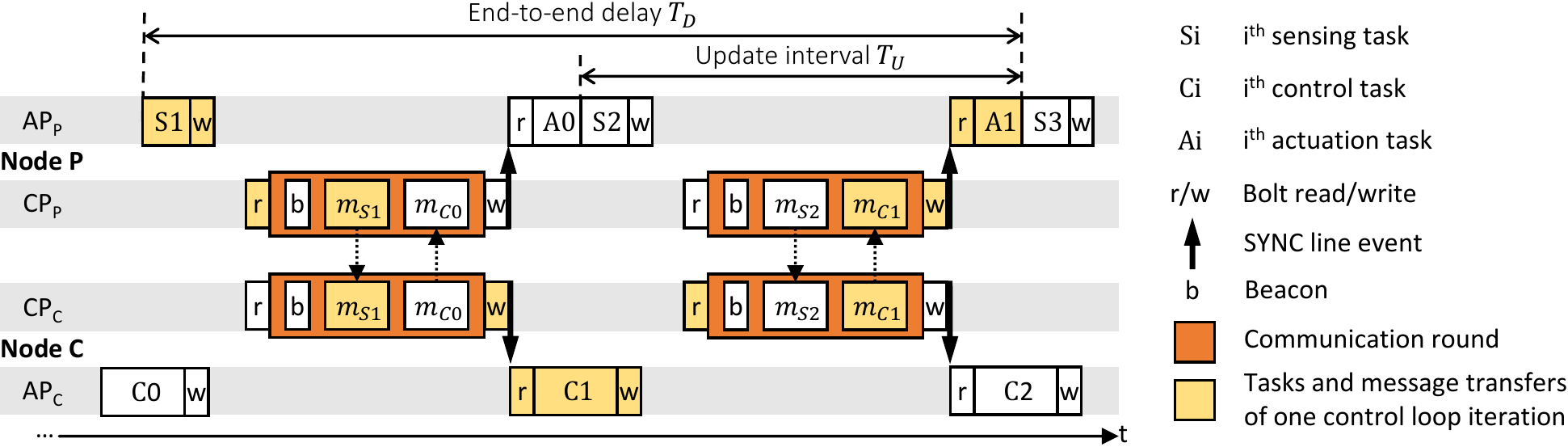}
	\caption{Example schedule of application tasks and message transfers between two dual-processor platforms. \capt{Node P senses and acts on a physical system, while node C runs the controller. In this example, the update interval \Tupdate is half the end-to-end delay \Tdelay.}} 
	\label{fig:schedule}
\end{figure*}

%

As shown in \figref{fig:protocol}, the operation of the protocol proceeds as a series of periodic \emph{communication rounds} with \emph{period}~$T$.
Each round consists of a sequence of non-overlapping time \emph{slots}.
In every time slot, all nodes in the network participate in a Glossy flood, where a message is sent from one node to all other nodes.
Glossy approaches the theoretical minimum latency for one-to-all flooding at a reliability above 99.9\percent, and provides microsecond-level network-wide time synchronization~\cite{Ferrari2011}.
In particular, a flood initiated by a dedicated node in the \emph{beacon} slot (b) at the beginning of every round is used for synchronization.
Nodes exploit the synchronization to remain in a low-power sleep mode between rounds and to awake in time for the next round, as specified by the round period~$T$.

It is important to note that, because of the way Glossy exploits \emph{synchronous transmissions}~\cite{Ferrari2011}, the wireless protocol operates \emph{independently} of the time-varying network topology.
In particular, this means that the wireless protocol and any logic built on top of it, such as a control algorithm, need not worry about the state of individual wireless links in the network.
This is a key difference to existing wireless protocols based on routing, such as WirelessHART and 6TiSCH, which, as we shall see, simplifies the control design and allows for providing formal guarantees that also hold in practice.


As detailed in \secref{subsec:scheduling}, we compute the communication schedule offline based on the traffic demands, and distribute it to all nodes before the application operation starts.
A schedule includes the assignment of messages to \emph{data} slots in each round (see \figref{fig:protocol}) and the round period~$T$.
Using static schedules brings several benefits.
First, we can a priori verify if closed-loop stability can be guaranteed for the achievable latencies (see \secref{sec:control}).
Second, compared to prior solutions~\cite{Ferrari2012,Istomin2016,Zimmerling2017,Jacob2016}, we can support significantly shorter latencies, and the protocol is more energy efficient (no need to send schedules) and more reliable (schedules cannot be lost).

%

\vsquish{-1mm}
\subsection{Hardware Platform}
\label{subsec:embeddedHardware}

\cps devices need to concurrently handle application tasks and message transfers.
While message transfers involve little but frequent computations, sensing and especially control tasks may require less frequent, but more demanding computations (\eg floating-point operations).
An effective approach to achieve low latency and high energy efficiency for such diverse workloads is to exploit hardware heterogeneity~(\textbf{G3}).

For this reason, we leverage a heterogeneous \emph{dual-processor platform (\dpp)}.
Application tasks execute exclusively on a 32-bit MSP432P401R ARM Cortex-M4F \emph{application processor} (\ap) running at 48\MHz, while the wireless protocol executes on a dedicated 16-bit CC430F5147 \emph{communication processor} (\cp) running at 13\MHz.
The \ap has a floating-point unit and a rich instruction set, accelerating computations related to sensing and control.
The \cp features a low-power microcontroller and a low-power wireless radio operating at 250\kbps in the 868\MHz frequency band.

\ap and \cp are interconnected using \bolt~\cite{Sutton2015a}, an ultra-low-power processor interconnect that supports asynchronous bidi\-rec\-tion\-al message passing with formally verified worst-case execution times.
\bolt decouples the two processors with respect to time, power, and clock domains, enabling energy-efficient concurrent executions with only small and bounded interference, thereby limiting jitter and preserving the time-sensitive operation of the wireless protocol.

All {\cp}s are time-synchronized via the wireless protocol.
In addition, \ap and \cp must be synchronized locally on each \dpp to minimize end-to-end delay and jitter among application tasks running on different {\ap}s~(\textbf{G1}).
To this end, we use a GPIO line between the two processors, called \emph{\sync} line.
Every \cp asserts the \sync line in response to an update of Glossy's time synchronization.
Every \ap schedules application tasks and message passing over \bolt with specific offsets relative to these \sync line events and resynchronizes its local time base.
Likewise, the {\cp}s execute the communication schedule and perform \sync line assertion and message passing over \bolt with specific offsets relative to the start of communication rounds.
As a result, all {\ap}s and {\cp}s act in concert.


\subsection{Scheduling Framework}
\label{subsec:scheduling}

We illustrate the scheduling problem we need to solve with a simple example, where
node P senses and acts on a physical system and node C runs the controller.

\figref{fig:schedule} shows a possible schedule of the application tasks and message transfers.
After sensing (S1), \apnode{P} writes a message containing the sensor reading into \bolt (w).
\cpnode{P} reads out the message (r) before the communication round in which that message ($m_{S1}$) is sent using the wireless protocol.
\cpnode{C} receives the message and writes it into \bolt.
After reading out the message from \bolt, \apnode{C} computes the control signal (C1) and writes a message containing it into \bolt.
The message ($m_{C1}$) is sent to \cpnode{P} in the next round, and then \apnode{P} applies the control signal on the physical system (A1).

This schedule resembles a pipelined execution, where in each communication round the last sensor reading and the next control signal (computed based on the previous sensor reading) are exchanged ($m_{S1} \, m_{C0}$, $m_{S2} \, m_{C1}$, $\ldots$).
Note that while it is indeed possible to send the corresponding control signal in the same round ($m_{S1} \, m_{C1}$, $m_{S2} \, m_{C2}$, $\ldots$), doing so would increase the update interval \Tupdate at least by the sum of the execution times of the control task, \bolt read, and \bolt write.
For the example schedule in \figref{fig:schedule}, the update interval \Tupdate is exactly half the end-to-end delay \Tdelay.

In general, the scheduling problem entails computing the communication schedule and the offsets with which all {\ap}s and {\cp}s in the system perform wireless communication, execute application tasks, transfer messages over \bolt, and assert the \sync line.
The problem gets extremely complex for any realistic scenario with more nodes or multiple feedback loops that are closed over the same wireless network, so solving it must be automated.

To this end, we leverage Time-Triggered Wireless~(\ttw)~\cite{Jacob2018}, an existing framework tailored to solve this type of scheduling problem.
\ttw takes as main input a dependency graph among application tasks and messages, similar to \figref{fig:overview}.
Based on an integer linear program, it computes the communication schedule and all offsets mentioned above.
\ttw provides three important guarantees: (\emph{i})~a feasible solution is found if one exists, (\emph{ii})~the solution minimizes the energy consumption for wireless communication, and (\emph{iii})~the solution can additionally optimize user-defined metrics (\eg minimize the update interval \Tupdate \mbox{as for the schedule in~\figref{fig:schedule}).}

\subsection{Essential Properties and Jitter Analysis}
\label{sec:properties}

%

\vspace{-1mm}
\fakepar{Essential properties}
The presented wireless embedded system design provides the following key properties for the control design:
\begin{itemize}
\item[\textbf{P1}] As analyzed below, for update intervals \Tupdate and end-to-end delays \Tdelay up to 100\ms, the worst-case jitter on \Tupdate and \Tdelay is bounded by \SI{\pm50}{\micro\second}.
It holds $\Tdelay = 2 \Tupdate$.
\item[\textbf{P2}] Statistical analysis of millions of Glossy floods~\cite{Zimmerling2013} and percolation theory for time-varying networks~\cite{Karschau2018} have shown that the spatio-temporal diversity in a Glossy flood reduces the temporal correlation in the series of received and lost messages by a node, to the extent that the series can be safely approximated by an \iid Bernoulli process.
The success probability in real multi-hop networks is typically larger than 99.9\percent~\cite{Ferrari2011}.
\item[\textbf{P3}] By provisioning for multi-hop many-to-all communication, arbitrary traffic patterns are efficiently supported.
\item[\textbf{P4}] It is guaranteed by design that message duplicates and out-of-order message deliveries do not occur.
\end{itemize}
%

\vspace{-1mm}
\fakepar{Worst-case jitter analysis}
To underpin \textbf{P1}, we analyze the \emph{worst-case} jitter on \Tupdate and \Tdelay.
We refer to \Tjitter as the nominal time interval between the end of two tasks executed on (possibly) different {\ap}s.
Due to jitter \Jitter, this interval may vary, resulting in an actual length of $\Tjitter + \Jitter$.
In our system, the jitter is bounded by
\begin{equation}
	|\, \Jitter \,| \leq 2 \, \left(\eREFmax + \eSYNCmax +  \Tjitter \, (\rhoAPmax + \rhoCPmax) \right) + \eTASKmax
	\label{eq:jitter_bound}
\end{equation}
where each term on the right-hand side of \eqref{eq:jitter_bound} is detailed below.


\emph{1) Time synchronization error between {\cp}s.}
Using Glossy, each \cp computes an estimate \trefEstim of the reference time~\cite{Ferrari2011} to schedule subsequent activities.
In doing so, each \cp makes an error \eREF with respect to the reference time of the initiator.
Using the approach from~\cite{Ferrari2011}, we measure \eREF for our Glossy implementation and a network diameter of up to nine hops.
Based on 340,000 data points, we find that \eREF ranges always between \SI{-7.1}{\micro\second} and \SI{8.6}{\micro\second}.
We thus consider $\eREFmax=\SI{10}{\micro\second}$ a safe bound for the jitter on the reference time between~{\cp}s.

\emph{2) Independent clocks on \cp and \ap.}
Each \ap schedules activities relative to \sync line events.
As \ap and \cp are sourced by independent clocks, it takes a variable amount of time until an \ap detects that \cp asserted the \sync line.
The resulting jitter is bounded by $\eSYNCmax=1/\fAP$, where $\fAP=48\MHz$ is the frequency of {\ap}s clock.

\emph{3) Different clock drift at {\cp}s and {\ap}s.}
The real offsets and durations of activities on the {\cp}s and {\ap}s depend on the frequency of their clocks.
Various factors contribute to different frequency drifts \rhoCP and \rhoAP, including the manufacturing process, ambient temperature, and aging effects.
State-of-the-art clocks, however, drift by at most $\rhoCPmax = \rhoAPmax = 50\ppm$~\cite{Lenzen2015}.

\emph{4) Varying task execution times.}
The difference between the task's best- and worst-case execution time $\eTASKmax$ adds to the jitter.
For the jitter on the update interval \Tupdate and the end-to-end delay \Tdelay, only the execution time of the actuation task matters, which typically exhibits little variance as it is short and highly deterministic.
For example, the actuation task in our experiments has a jitter of $\SI{\pm3.4}{\micro\second}$.
To be safe, we consider $\eTASKmax = \SI{10}{\micro\second}$ for our analysis.

Using \eqref{eq:jitter_bound} and the above values, we can compute the worst-case jitter for a given interval \Tjitter.
Fast feedback control as considered in this paper requires $\Tjitter = \Tdelay = 2 \Tupdate \leq 100\ms$, which gives a worst-case jitter of \SI{\pm50}{\micro\second} on \Tupdate and \Tdelay, as stated by property \textbf{P1}.


\section{Control Design and Analysis}
\label{sec:control}

Building on the design of the wireless embedded system and its properties \textbf{P1}--\textbf{P4}, this section addresses the design of the control system to accomplish goals \textbf{G4}--\textbf{G6} from \secref{sec:overview}.
Because the wireless system supports arbitrary traffic patterns (\textbf{P3}), various control tasks can be solved regardless of whether sensors, actuators, physical system(s), and controller(s) are co-located or spatially distributed.
This includes typical single-loop tasks such as stabilization, disturbance rejection, or set-point tracking, as well as multi-agent scenarios such as synchronization, consensus, or formation control.


Here, we focus on remote stabilization and synchronization of multiple agents over wireless multi-hop networks as prototypical examples for both the single- and multi-agent case.
For stabilization, modeling and control design are presented in Sections~\ref{sec:model} and \ref{sec:ctrlDesign}, thus achieving \textbf{G5}.
The stability analysis is provided in \secref{sec:stabAnalysis}, which fulfills \textbf{G4}.
Synchronization is discussed in \secref{sec:sync}, highlighting support for straightforward distributed control \textbf{G6}.

\subsection{Model of Wireless Control System}
\label{sec:model}
\begin{figure}[!tb]
\centering
\tikzsetnextfilename{wirelessControlSystemModel}
\tikzset{radiation/.style={{decorate,decoration={expanding
waves,angle=90,segment length=4pt}}}}
\tikzset{>=latex}
\tikzset{font=\scriptsize}
\begin{tikzpicture}
\node[draw,rectangle, minimum height = 2em,align=center,minimum width = 5em](controller){Controller\\ $\hat{x}(k)$};
.\node[draw, rectangle, minimum width = 7.5em,above = 2.25em of controller, minimum height = 1em](network){Wireless Network};
\node[draw,rectangle,minimum height = 2.5em,above = 2.75em of network,align = center](system){Physical System\\$x(k)$};
\node[draw,rectangle,below=0em of system.south west, anchor = north west,inner sep=0,minimum width=0.8em, minimum height=0.8em](act){A};
\node[draw,rectangle,below=0em of system.south east, anchor = north east,inner sep=0,minimum width=0.8em, minimum height=0.8em](sens){S};
\draw[->] (controller.north-|act)-- node[midway,right]{$\hat{u}(k+1)$} (network.south-|act);
\draw[->] (network.south-|sens)-- node[midway,right, align = center]{$y(k-1)$\\ if $\theta\!=\!1$} (controller.north-|sens);
\draw[->] (network.north-|act)-- node[midway,right,align=center]{$\hat{u}(k)$\\if $\phi\!=\!1$} (act.south);
\draw[->] (sens.south)-- node[midway,right]{$y(k)$} (network.north-|sens);
\end{tikzpicture}
 \tikzsetnextfilename{wirelessSyncSystemModel}
\begin{tikzpicture}
\node[draw,rectangle,minimum height = 2.5em,align = center](system1){Physical\\ System 1\\$x_1(k)$};
\node[draw,rectangle,below=0em of system1.north west, anchor = south west,inner sep=0,minimum width=0.8em, minimum height=0.8em](act1){A};
\node[draw,rectangle,below=0em of system1.north east, anchor = south east,inner sep=0,minimum width=0.8em, minimum height=0.8em](sens1){S};
\node[draw,rectangle, minimum height = 2em,align=center, right = 1.5em of system1](system2){Physical\\ System 2\\ $x_2(k)$};
\node[draw,rectangle,below=0em of system2.north west, anchor = south west,inner sep=0,minimum width=0.8em, minimum height=0.8em](sens2){S};
\node[draw,rectangle,below=0em of system2.north east, anchor = south east,inner sep=0,minimum width=0.8em, minimum height=0.8em](act2){A};
\node[draw, rectangle, minimum width = 10em,above = 8em of $(system1)!0.5!(system2)$, minimum height = 1em](network){Wireless Network};
\node[draw, rectangle, above = 2em of act1](ctrl1){Ctrl 1};
\node[draw,rectangle,above=2em of act2](ctrl2){Ctrl 2};
\draw[->] (sens2)-- node[pos=0.18,right]{$y_2(k)$} (network.south-|sens2);
\draw[->](sens2) |- (ctrl2);
\draw[->] (network.south-|act2)-- node[midway,right,align=center]{$ y_1(k-1)$\\if $\phi\!=\!1$} (ctrl2);
\draw[->](ctrl1) -- node[midway, left]{$u_1(k)$} (act1);
\draw[->] (network.south-|act1)-- node[midway,left,align=center]{$y_2(k-1)$\\if $\theta\!=\!1$} (ctrl1);
\draw[->](ctrl2) -- node[midway,right]{$u_2(k)$}(act2);
\draw[->] (sens1)-- node[pos=0.18,left]{$y_1(k)$} (network.south-|sens1);
\draw[->](sens1) |- (ctrl1);
\end{tikzpicture}
%
\caption{
Wireless control tasks: stabilization (left) and synchronization (right).
\capt{The feedback loop for stabilizing the physical system (left) is closed over the multi-hop low-power wireless network. This induces delays and message loss, which is captured by \iid Bernoulli variables $\theta$ and $\phi$. Two physical systems, each with a local controller (Ctrl), are synchronized over the wireless network (right).}
}
\label{fig:WirelessControlModel}
\end{figure}
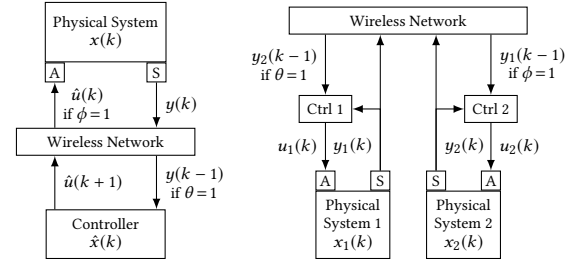

We address the remote stabilization task depicted in \figref{fig:WirelessControlModel} (left), where controller and physical system are associated with different nodes, which can communicate via the wireless network.
Such a scenario is relevant for instance in process control, where the controller often resides at a remote location~\cite{Ma2018}.
We consider stochastic LTI dynamics for the physical process
\begin{subequations}
\label{eqn:sys_complete}
\begin{align}
\label{eqn:ssmodel}
x(k+1) = Ax(k) + Bu(k) + v(k).
\end{align}
This model describes the evolution of the system state $x(k)\in\mathbb{R}^n$ with discrete time index $k \in \mathbb{N}$ in response to the \mbox{control input} \mbox{$u(k)\in\mathbb{R}^m$} and random process noise $v(k)\in\mathbb{R}^n$.
As typical in the literature~\cite{Astrom2008,Hespanha2007}, the process noise is modeled as an \iid Gaussian random variable with zero mean and variance $\Sigma_\mathrm{proc}$, \mbox{$v(k)\sim \mathcal{N}(0,\Sigma_\mathrm{proc})$}, and captures, for example, uncertainty in the model.

We assume that the full system state $x(k)$ can be measured through appropriate sensors, that is,
\begin{align}
\label{eqn:meas}
y(k) = x(k)+w(k),
\end{align}
\end{subequations}
with sensor measurements $y(k)\in\mathbb{R}^n$ and sensor noise $w(k)\in\mathbb{R}^n$, $w(k)\sim \mathcal{N}(0,\Sigma_\mathrm{meas})$.
If the complete state vector cannot be measured directly, it can typically be reconstructed via state estimation techniques \cite{Astrom2008}.

The process model in \eqref{eqn:sys_complete} is stated in discrete time. 
This representation is particularly suitable here as the wireless embedded system offers a constant update interval \Tupdate with worst case jitter of \SI{\pm50}{\micro\second} (\textbf{P1}), which can be neglected from controls perspective~\cite[p.~48]{Cervin2003}.
Thus, $u(k)$ and $y(k)$ in \eqref{eqn:sys_complete} represent sensing and actuation at periodic intervals \Tupdate, as illustrated in \figref{fig:schedule}.

As shown in \figref{fig:WirelessControlModel}, measurements $y(k)$ and control inputs $\hat{u}(k)$ are sent over the wireless network.
According to \textbf{P1} and \textbf{P2}, both arrive at the controller, respectively system, with a delay of \Tupdate and with a probability governed by two independent Bernoulli processes.
We represent the Bernoulli processes by $\theta(k)$ and $\phi(k)$, which are \iid binary variables, indicating lost ($\theta(k) = 0$, $\phi(k)=0$) or successfully received ($\theta(k)=1$, $\phi(k)=1$) messages.
To ease notation and since both variables are \iid, we can omit the time index in the following without any confusion.
We denote the probability of successful message delivery by $\mu_\theta$ (\ie $\mathbb{P}[\theta=1] = \mu_\theta$), respectively $\mu_\phi$.
As both, measurements and control inputs, are delayed, it also follows that in case of no message losses, the applied control input $u(k)$ depends on the measurement two steps ago $y(k-2)$.
If a control input message is lost, the input stays constant since zero-order hold is used at the actuator, that is,
\begin{align}
\label{eqn:ZOH}
u(k)=\phi \hat{u}(k)+\left(1-\phi\right)u(k-1).
\end{align}

The model proposed in this section thus captures the properties \textbf{P1}, \textbf{P2}, and \textbf{P4}.
While \textbf{P1} and \textbf{P2} are incorporated in the presented dynamics and message loss models, \textbf{P4} means that there is no need to take duplicated or out-of-order sensor measurements and control inputs into account.
Overall, these properties allow for accurately describing the wireless \cps by a fairly straightforward model, which greatly facilitates subsequent control design and analysis.
Property \textbf{P3} is not considered here, where we deal with a single control loop, but will  become essential in \secref{sec:sync}.

\vsquish{-2mm}
\subsection{Controller Design}
\label{sec:ctrlDesign}
Designing a feedback controller for the system in \eqref{eqn:sys_complete}, we proceed by first discussing state-feedback control for the nominal system (\ie without delays, message loss, and noise), and then enhance the design to cope with the network and sensing imperfections.

\fakepar{Nominal design}
Assuming ideal measurements, \mbox{$y(k)\! =\! x(k)$} holds.
A common strategy in this setting is static state-feedback control, $u(k)\! =\! Fx(k)$, where $F$ is a constant feedback matrix, which can be designed, for instance, via \emph{pole placement} or methods from optimal control, such as the \emph{linear quadratic regulator (LQR)} \cite{Astrom2008,Anderson2007}.
Under the assumption of controllability~\cite{Astrom2008}, desired (in particular, stable) dynamics can be obtained for the state in \eqref{eqn:ssmodel} in this way.

\fakepar{Actual design}
We augment the nominal state-feedback design to cope with non-idealities, in particular, delayed measurements and message loss, as shown in \figref{fig:WirelessControlModel} (left).

Because the measurement arriving at the controller $y(k-1)$ represents information that is one time step in the past, the controller propagates the system for one step as follows:
\begin{align}
\hat{x}(k) &= \theta Ay(k\!-\!1)+(1\!-\!\theta)(A\hat{x}(k\!-\!1))+B\hat{u}(k\!-\!1) \label{eqn:prediction} \\
&= \theta Ax(k\!-\!1) + (1\!-\!\theta)A\hat{x}(k\!-\!1)+B\hat{u}(k\!-\!1)+\theta Aw(k\!-\!1), \nonumber
\end{align}
where $\hat{x}(k)$ is the predicted state, and $\hat{u}(k)$ is the control input
computed by the controller
(to be made precise below).  Both variables are computed by the controller and represent its internal states.  The rationale of \eqref{eqn:prediction} is as follows:  If the measurement message is delivered (the controller has information about $\theta$ because it knows when to expect a message), we compute the state prediction based on this measurement $y(k\!-\!1)\! = \!x(k\!-\!1)+w(k\!-\!1)$; if the message is lost, we propagate the previous prediction $\hat{x}(k\!-\!1)$.
As there is no feedback on lost control messages (\ie about $\phi$) and thus a potential mismatch between the computed input $\hat{u}(k\!-\!1)$ and the actual $u(k\!-\!1)$, the controller can only use $\hat{u}(k\!-\!1)$
in the prediction.

Using $\hat{x}(k)$, the controller has an estimate of the current state of the system.
However, it will take another time step for the currently computed control input to arrive at the physical system.  For computing the next control input, we thus propagate the system another step,
\begin{align}
\label{eqn:pred_inp}
\hat{u}(k+1)&=F\left(A\hat{x}(k)+B\hat{u}(k)\right),
\end{align}
where $F$ is as in the nominal design.
The input $\hat{u}(k+1)$ is then transmitted over the wireless network (see \figref{fig:WirelessControlModel}, left).

The overall controller design requires only a few matrix multiplications per execution.
This can be efficiently implemented on embedded devices, thus satisfying goal \textbf{G5}.

\subsection{Stability Analysis}
\label{sec:stabAnalysis}
We now present a stability proof for the closed-loop system given by the dynamic system described in \secref{sec:model}
and the proposed controller from \secref{sec:ctrlDesign}.  Because the model in \secref{sec:model} incorporates the physical process and the essential properties of the wireless embedded system, we achieve goal \textbf{G4}.

While the process dynamics in \eqref{eqn:sys_complete} are time invariant, message loss introduces time variation and randomness into the system dynamics.
Therefore, we leverage stability results for linear, stochastic,
time-varying systems~\cite{Boyd1994}.
For ease of exposition, we consider \eqref{eqn:sys_complete} without process and measurement noise (\ie $v(k)=0$ and $w(k)=0$), and comment later on extensions.
 We first introduce required definitions and preliminary results, and then apply those results to our problem.


Consider the system
\begin{align}
\label{eqn:gen_sys}
z(k+1)=\tilde{A}(k)z(k)
\end{align}
with state $z(k)\in\mathbb{R}^n$
and $\tilde{A}(k) = \tilde{A}_0 + \sum_{i=1}^L \tilde{A}_ip_i(k)$; the $p_i(k)$ are \iid random variables with mean $\E[p_i(k)]=0$, variance \mbox{$\Var[p_i(k)]=\sigma_{p_i}^2$}, and $\E[p_i(k)p_j(k)]=0 \,\forall i,j$.

A common notion of stability for stochastic systems like the one in \eqref{eqn:gen_sys} is mean-square stability:
\begin{defi}[{\cite[p.~131]{Boyd1994}}]
Let $Z(k) := \E[z(k)z^\mathrm{T}(k)]$ denote the state correlation matrix. The system in \eqref{eqn:gen_sys} is \emph{mean-square stable (MSS)} if $\,\lim_{k\to\infty} Z(k) = 0$
for any initial $z(0)$.
%
\end{defi}
\noindent
That is, a system is called MSS if the state correlation vanishes asymptotically for any intial state.  MSS implies, for example, that $z(k) \to 0$ almost surely as $k\to\infty$~\cite[p.~131]{Boyd1994}.

In control theory, linear matrix inequalities (LMIs) are often used as computational tools to check for system properties such as stability (see \cite{Boyd1994} for an introduction and details).  For MSS, we employ the following LMI stability result:
\begin{lem}[{\cite[p.~131]{Boyd1994}}]
\label{lem:LMIcond}
System~\eqref{eqn:gen_sys} is MSS if, and only if, there exists a positive definite matrix
$P>0$ such that
\begin{equation}
\label{eqn:LMI}
\tilde{A}_0^\mathrm{T}P\tilde{A}_0-P+\sum\nolimits_{i=1}^N\sigma_{p_i}^2\tilde{A}^\mathrm{T}_iP\tilde{A}_i <0 .
\end{equation}
\end{lem}

We now apply this result to the system and controller from Sections \ref{sec:model} and \ref{sec:ctrlDesign}.
The closed-loop dynamics are given by \eqref{eqn:sys_complete}--\eqref{eqn:pred_inp}, which we rewrite
%
as an augmented system
\begin{align}
\label{eqn:matrix_repr}
\underbrace{\begin{pmatrix}
x(k+1)\\\hat{x}(k+1)\\u(k+1)\\ \hat{u}(k+1)
\end{pmatrix}}_{z(k+1)}
&=\underbrace{\begin{pmatrix}
A&0&B&0\\
\theta A&(1-\theta)A&0&B\\
0&\phi FA&(1-\phi)I&\phi FB\\
0&FA&0&FB
\end{pmatrix}}_{\tilde{A}(k)}
\underbrace{\begin{pmatrix}
x(k)\\ \hat{x}(k)\\u(k)\\ \hat{u}(k)
\end{pmatrix}}_{z(k)}.
\end{align}
The system has the form of \eqref{eqn:gen_sys}; the transition matrix depends on $\theta$ and $\phi$, and thus on time (omitted for simplicity).  We can thus apply Lemma~\ref{lem:LMIcond} to obtain our main stability result, whose proof is given in \appref{sec:app_proof}.
\begin{theo}
\label{thm:MSSourSystem}
The system~\eqref{eqn:matrix_repr} is MSS if, and only if, there exists a $P>0$ such that \eqref{eqn:LMI} holds with
\begin{align*}
\tilde{A}_0 &= \left(\begin{smallmatrix}
A&0&B&0\\
\mu_\theta A&(1-\mu_\theta)A&0&B\\
0&\mu_\phi FA&(1-\mu_\phi)I&\mu_\phi FB\\
0&FA&0&FB
\end{smallmatrix}\right), \quad
\tilde{A}_1 = \left(\begin{smallmatrix}
0&0&0&0\\
-\mu_\theta A&\mu_\theta A&0&0\\
0&0&0&0\\
0&0&0&0
\end{smallmatrix}\right), \\
\tilde{A}_2 &= \left(\begin{smallmatrix}
0&0&0&0\\
0&0&0&0\\
0&-\mu_\phi FA&\mu_\phi I & -\mu_\phi FB\\
0&0&0&0
\end{smallmatrix}\right), \quad
\sigma_{p_1}^2 = \sfrac{1}{\mu_\theta}-1, \quad
\sigma_{p_2}^2 = \sfrac{1}{\mu_\phi} -1.
\end{align*}
\end{theo}

Using Theorem~\ref{thm:MSSourSystem}, we can analyze stability for any concrete physical system
\eqref{eqn:sys_complete},
a state-feedback controller $F$, and probabilities $\mu_\theta$ and $\mu_\phi$.
Searching for a $P>0$ that satisfies the LMI in~\eqref{eqn:LMI} can be done using efficient numerical tools based on convex optimization (\eg \cite{Labit2002}). 
If such a $P$ is found, we have the stability guarantee~(\textbf{G4}).

The stability analysis can be extended to account for process and measurement noise so that MSS then implies bounded $Z(k)$ (see \cite[p.~138]{Boyd1994}).
Moreover, other combinations of end-to-end delay \Tdelay and update interval \Tupdate are possible, including $\Tdelay = n\Tupdate$ ($n\in\mathbb{N}$).
Also the sensor-to-controller and controller-to-actuator delays may be different.

\subsection{Multi-agent Synchronization}
\label{sec:sync}
In distributed or decentralized control architectures, different controllers have access to different measurements and inputs, and thus, in general, different information.
This is the core reason for why such architectures are more challenging than centralized ones \cite{Lunze1992,Grover2014}.
Which information a controller has access to depends on the traffic pattern and topology of the network.
For instance, an agent may only be able to communicate with its nearest neighbor via point-to-point communication,  	
or with other agents in a certain range. 
Property \textbf{P3} of the wireless embedded system in \secref{sec:embedded} offers a key advantage compared to these structures because every agent in the network has access to all information (except for rare message losses).
We can thus carry out a centralized design, but implement the resulting controllers in a distributed fashion (cf. \figref{fig:WirelessControlModel}, right).
Such schemes have been used before for wired-bus networks (\eg in \cite{Trimpe2012}).

Here, we present synchronization of multiple physical systems as an \emph{example} of how distributed control tasks can easily be achieved with the proposed wireless control system (\textbf{G6}).
We assume multiple physical processes as in~\eqref{eqn:sys_complete}, but with possibly different dynamics parameters ($A_i$, $B_i$, \etc).
We understand synchronization in this setting as the goal of having the system state of different agents evolve together as close as possible.
That is, we want to keep the error $x_i(k)-x_j(k)$ between the states of systems $i$ and $j$ small.
Instead of synchronizing the whole state vector, also a subset of all states can be considered.
Synchronization of multi-agent systems is a common problem and also known as consensus or coordination~\cite{Lunze2012}.

We demonstrate feasibility of synchronization with multiple systems in \secref{sec:synchronization}.
The synchronizing controller is based on an LQR~\cite{Anderson2007}; details of the concrete design are given in \appref{sec:app_sync}.
\section{Experimental Evaluation}
\label{sec:eval}

This section uses measurements from a cyber-physical testbed (see \figref{fig:testbed}) consisting of 20 wireless embedded devices (forming a three-hop network) and several cart-pole systems to evaluate the performance of the proposed wireless \cps design.
Our experiments reveal the following key findings:
\begin{itemize}
 \item We can safely stabilize two inverted pendulums via two remote controllers across the three-hop low-power wireless network.
 \item Using the same \cps design with a different control logic, we can reliably synchronize the movement of five cart-poles thanks to the support for arbitrary traffic patterns.
 \item Our system can stabilize an inverted pendulum at update intervals of 20--50\ms. Increasing the update interval decreases the control performance, but leads to significant energy savings on the wireless communication side.
 \item Our system is highly robust to message loss. Specifically, it can stabilize an inverted pendulum at an update interval of \SI{20}{\ms} despite 75\% \iid Bernoulli losses and in situations with bursts of 40 consecutively lost messages.
 \item The measured jitter on the update interval and the end-to-end delay is less than $\SI{\pm25}{\micro\second}$, which validates our analysis of the theoretical worst-case jitter of $\SI{\pm50}{\micro\second}$ from \secref{sec:properties}.
\end{itemize}


\vsquish{-2mm}
\subsection{Cyber-physical Testbed}
\label{sec:cps_testbed}

\begin{figure}[!tb]
	\centering
	\includegraphics[width=0.95\linewidth]{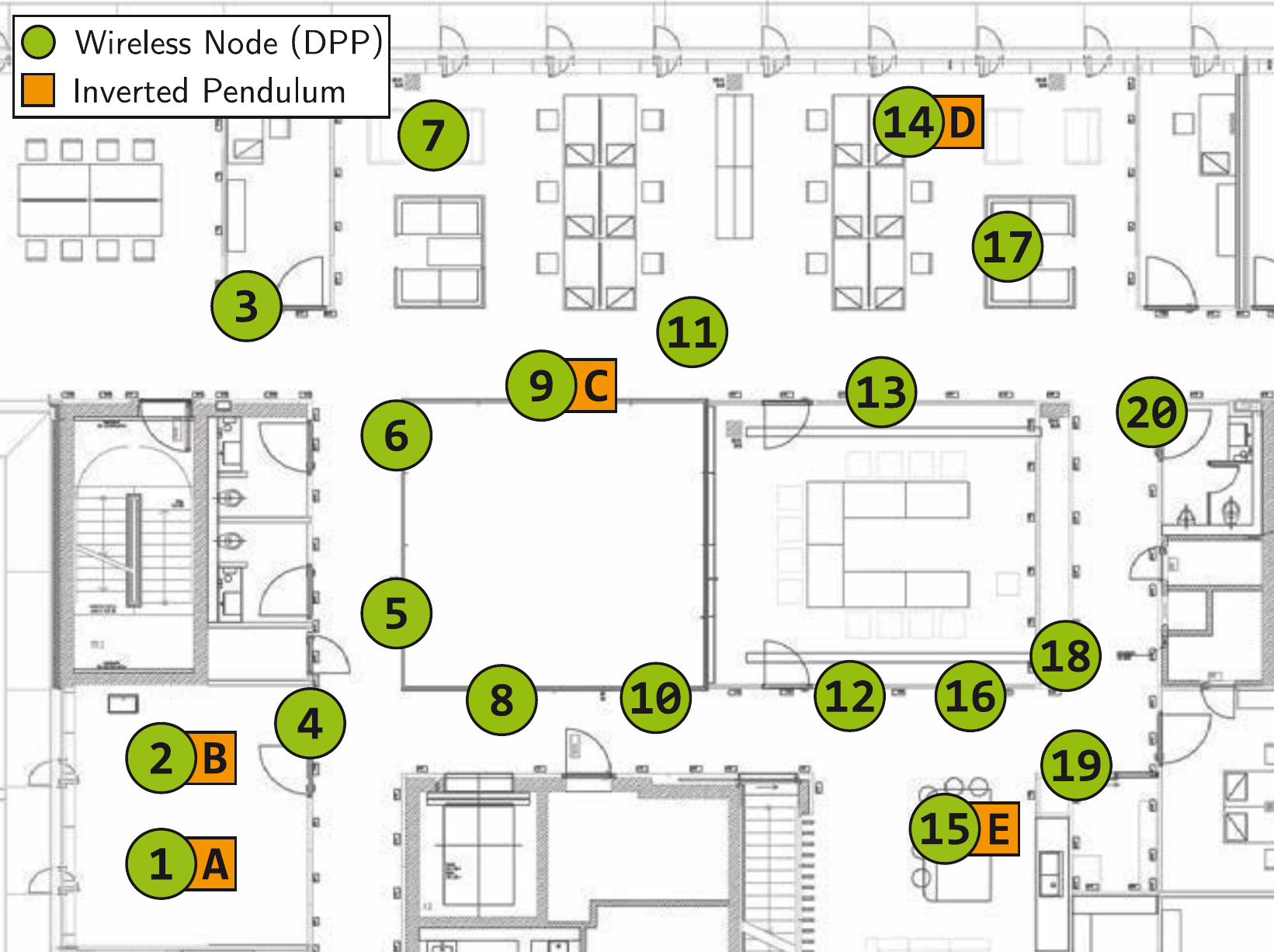}
	\vsquish{-3mm}
	\caption{Layout of cyber-physical testbed consisting of 20 \dpp nodes that form a three-hop low-power wireless network and five cart-pole systems (two real ones attached to nodes 1 and 2, three simulated ones at nodes 9, 14, and 15).}
	\vsquish{-3mm}
	\label{fig:testbed}
\end{figure}

Realistic cyber-physical testbeds are essential for the validation and evaluation of \cps solutions~\cite{Lu2016,Baumann2018}.
We developed the wireless cyber-physical testbed depicted in \figref{fig:testbed}.
It consists of 20 \dpp nodes, two real physical systems (A and B), and three simulated physical systems (C, D, and E).
The testbed is deployed in an office building and extends across an area of \SI{15}{\m} by \SI{20}{\m}.
All nodes transmit at \SI{10}{\dBm}, which results in a network diameter of three hops.
The wireless signals need to penetrate various types of walls, from glass to reinforced concrete, and are exposed to different sources of interference from other electronics and human activity.

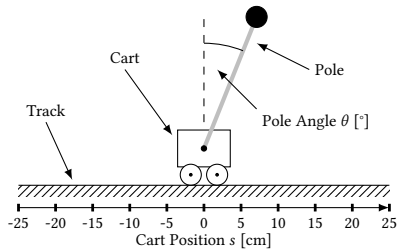
\begin{figure}[!tb]
\centering
\tikzsetnextfilename{cartPole}
\begin{tikzpicture}[scale=0.7, every node/.style={scale=0.7}]]
\tikzstyle{ground}=[fill,pattern=north east lines,draw=none,minimum width=5,minimum height=0.1]
\tikzset{>=latex};
\draw(0.5,0.35)rectangle(-0.5,-0.35);
\draw[fill=white](-0.25,-0.5)circle(0.2);
\draw[fill=white](0.25,-0.5)circle(0.2);
\draw[ultra thick,lightgray](0,0)node(cart){}--(1,2.5);
\draw[fill=black](1,2.5)circle(0.2)node(pole){};
\draw[dashed] (0,0.5) -- (0,2.5)node(top){};
\draw[fill=black](0,0)circle(0.05);
\draw[fill=black](-0.25,-0.5)circle(0.025);
\draw[fill=black](0.25,-0.5)circle(0.025);
\node[ground,minimum width=200,anchor=north](floor)at(0,-0.7){};
\draw(floor.north east)--(floor.north west);
 \draw pic[" ",draw=black, -,  angle eccentricity=1.5,angle radius =
  2cm] {angle = pole--cart--top};
\draw[->](-1.5,1.5)node[above]{Cart} -- (-0.6,0.25);
\draw[->] (2,1.5)node[right]{Pole} -- (1,2);
\draw[->]([shift={(0,-0.2cm)}]floor.south west) -- node[midway,draw,rectangle,inner sep = 0, minimum width=0.1em, minimum height=0.4em,label=below:0,fill=black]{}
node[pos=0.4,draw,rectangle,inner sep = 0, minimum width=0.1em, minimum height=0.4em,label=below:-5,fill=black]{}
node[pos=0.3,draw,rectangle,inner sep = 0, minimum width=0.1em, minimum height=0.4em,label=below:-10,fill=black]{}
node[pos=0.2,draw,rectangle,inner sep = 0, minimum width=0.1em, minimum height=0.4em,label=below:-15,fill=black]{}
node[pos=0.1,draw,rectangle,inner sep = 0, minimum width=0.1em, minimum height=0.4em,label=below:-20,fill=black]{}
node[pos=0,draw,rectangle,inner sep = 0, minimum width=0.1em, minimum height=0.4em,label=below:-25,fill=black]{}
node[pos=0.6,draw,rectangle,inner sep = 0, minimum width=0.1em, minimum height=0.4em,label=below:5,fill=black]{}
node[pos=0.7,draw,rectangle,inner sep = 0, minimum width=0.1em, minimum height=0.4em,label=below:10,fill=black]{}
node[pos=0.8,draw,rectangle,inner sep = 0, minimum width=0.1em, minimum height=0.4em,label=below:15,fill=black]{}
node[pos=0.9,draw,rectangle,inner sep = 0, minimum width=0.1em, minimum height=0.4em,label=below:20,fill=black]{}
node[pos=1,draw,rectangle,inner sep = 0, minimum width=0.1em, minimum height=0.4em,label=below:25,fill=black]{}
([shift={(0,-0.2cm)}]floor.south east);
\draw[->] (-3,0.5)node[above]{Track} -- ([shift={(-2.5,0.1cm)}]floor.north);
\node[below = 1.5em of floor,inner sep = 0]{Cart Position $s$ [\si{\centi\meter}]};
\draw[->](1,0.75)node[below right]{Pole Angle $\theta$ [\si{\degree}]} -- (0.25,1.5);
\end{tikzpicture}
%
\vsquish{-2mm}
\caption{Schematic of a cart-pole system used in our testbed as physical systems. \capt{By controlling the force applied to
the cart, the pole can be stabilized in the upright position around $\theta=\SI{0}{\degree}$.}}
\vsquish{-8mm}
\label{fig:pendulum}
\end{figure}


We use \emph{cart-pole systems} as physical systems.
As shown in \figref{fig:pendulum}, a cart-pole system consists of a cart that can move horizontally on a track and a pole attached to it via a revolute joint.
The cart is equipped with a DC motor that can be controlled by applying a voltage to influence the speed and the direction of the cart.
Moving the cart exerts a force on the pole and thus influences the pole angle~$\theta$.
This way, the pole can be stabilized in an upright position around $\theta=\SI{0}{\degree}$, which represents an unstable equilibrium and is called the \emph{inverted pendulum}.
The inverted pendulum has fast dynamics, which are typical of real-world mechanical systems~\cite{Boubaker2012}, and requires feedback with update intervals of tens of milliseconds.

%

For small deviations from the equilibrium (\ie $\sin(\theta) \approx \theta$), the inverted pendulum can be well approximated by an LTI system.
The state $x(k)$ of the system consists of four variables.
Two of them, the pole angle~$\theta(k)$ and the cart position~$s(k)$, are measured by angle sensors.
Their derivatives, the angular velocity~$\dot{\theta}(k)$ and the cart velocity~$\dot{s}(k)$, are estimated using finite differences and low-pass filtering.
The voltage applied to the motor is the control input $u(k)$.
In this way, the APs of nodes 1 and 2 interact with the two real pendulums A and B, while the APs of nodes 9, 14, and 15 run simulation models of the inverted pendulum.

The cart-pole system has a few constraints.
Control inputs are capped at \SI{\pm10}{\V}.
The track has a usable length of \SI{\pm25}{\cm} from the center (see \figref{fig:pendulum}).
Surpassing the track limits immediately ends an experiment.
At the beginning of an experiment, we move the carts to the center and the poles in the upright position; then the controller takes over.
Appendices~\ref{sec:app_stab_ctrl} and~\ref{sec:app_sync} detail the implementation of the controllers for multi-hop stabilization and multi-hop synchronization, following the design outlined in \mbox{Sections~\ref{sec:ctrlDesign} and~\ref{sec:sync}}.

Using this cyber-physical testbed, we measure the control performance in terms of pole angle, cart position, and control input.
In addition, we measure the radio duty cycle at each node in software and record messages that are lost over the wireless network.


\vsquish{-5mm}
\subsection{Multi-hop Stabilization}
\label{sec:multihop_stabilization}

In our first experiment, we want to answer the main question of this work and investigate the feasibility of fast feedback control over multi-hop low-power wireless networks.

\fakepar{Setup}
We use two controllers running on nodes 14 and 15 to stabilize the two real pendulums A and B at $\theta = \SI{0}{\degree}$ and $s = \SI{0}{\cm}$.
Hence, there are two independent control loops sharing the same wireless network, and it takes in total six hops to close each loop.
We configure the wireless embedded system and the controllers for an update interval of $\Tupdate = \SI{45}{\ms}$.
As per property~\textbf{P2} and confirmed by our measurements discussed below, we expect a message delivery rate of at least 99.9\percent.
With this we have $\mu_\theta = \mu_\phi = 0.999$, and we can prove stability of the overall system using Theorem~\ref{thm:MSSourSystem}.



\begin{figure}[!tb]
	\centering
	\includegraphics[width=\linewidth]{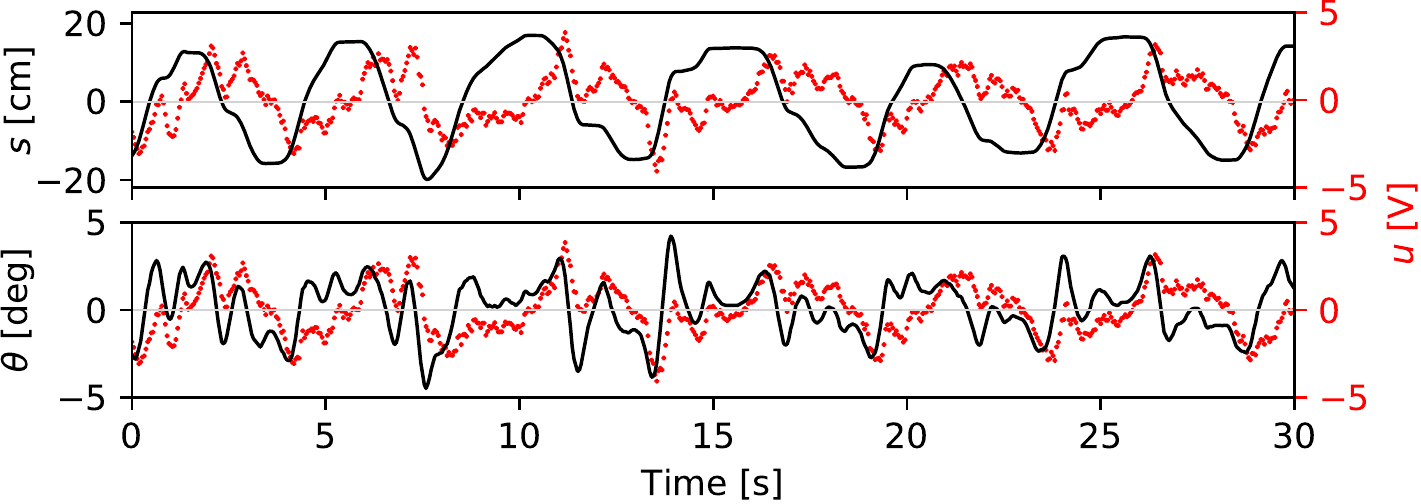}
	\vspace{-5mm}
	\caption{Cart position \boldmath$s$, pole angle \boldmath$\theta$, and control input \boldmath$u$ of one cart-pole system when concurrently stabilizing two cart-pole systems over a multi-hop network at an update interval of 45\ms. \capt{The cart position and the pole angle always stay within safe regimes, and less than half of the possible control input is needed.}}
	\vsquish{-5mm}
	\label{fig:stabilization_angle_voltage_time_plot}
\end{figure}

\begin{figure}[!tb]
	\centering
	\includegraphics[width=\linewidth]{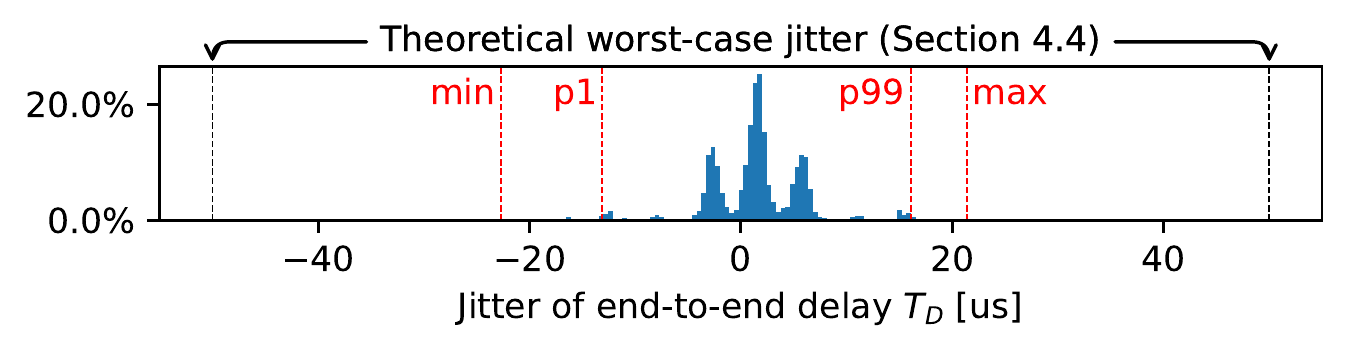}
	\vspace{-6mm}
	\caption{Distribution of the jitter on the end-to-end delay \boldmath$\Tdelay$; results for the update interval \boldmath$\Tupdate$ are similar.  \capt{The measurements are within the worst-case bounds determined in \secref{sec:properties}.}}
	\vspace{-4mm}
	\label{fig:update_interval_jitter}
\end{figure}

\fakepar{Results}
The experimental results confirm the theoretical analysis: We are able to safely stabilize both pendulums over the three-hop wireless network, even while carrying around the controller because our design is independent of the network topology~\footnote{A video of this experiment can be found at \url{https://youtu.be/19xPHjnobkY}.} (see \secref{sec:wireless_protocol}).
\figref{fig:stabilization_angle_voltage_time_plot} shows a characteristic \SI{30}{\second} trace of one of the pendulums.
Cart position, pole angle, and control input oscillate, but always stay within safe regimes.
For example, the cart never comes close to either end of the track and less than half of the possible control input is needed to stabilize the pendulum.
Not a single message was lost in this experiment, which demonstrates the reliability of our wireless embedded system design. 

During the same experiment, we also use a logic analyzer to continuously measure the update interval \Tupdate and the end-to-end delay \Tdelay (see \figref{fig:schedule}).
\figref{fig:update_interval_jitter} shows the measured jitter on \Tdelay; the results for \Tupdate look very similar.
We see that the empirical results are well within the theoretical worst-case bounds, which validates our analysis in \secref{sec:properties} and assumptions in \secref{sec:control}.

\vsquish{-2mm}
\subsection{Multi-hop Synchronization}
\label{sec:synchronization}

We now apply the same wireless \cps design to a distributed control task to demonstrate its flexibility and versatility.

\fakepar{Setup}
We use the two real pendulums A and B and the three simulated pendulums C, D, and E.
The goal is to synchronize the cart positions of the five pendulums over the wireless multi-hop network, while each pendulum is stabilized  by a local control loop.
This scenario is similar to drone swarm coordination, where each drone stabilizes its flight locally, but exchanges its position with all other drones to keep a desired swarm formation~\cite{Preiss2017}.
In our experiment, stabilization runs with $\Tupdate = \SI{10}{\milli\second}$, and nodes 1, 2, 9, 14, and 15 exchange their current cart positions every \SI{50}{\milli\second}.

\begin{figure*}[!bt]
	\centering
	\includegraphics[width=\linewidth]{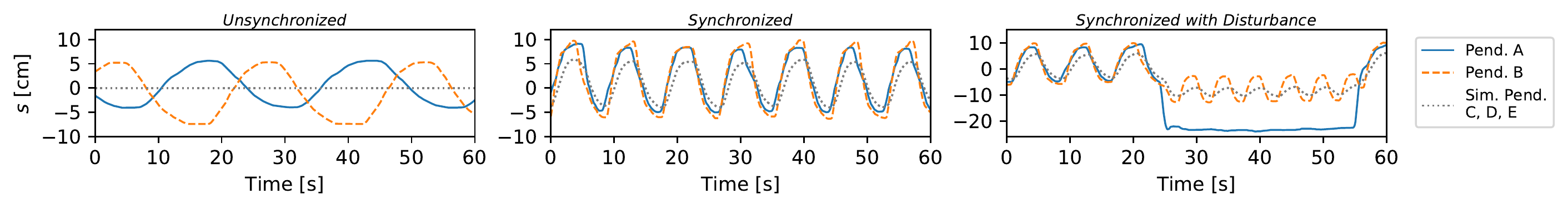}
	\vspace{-8mm}
	\caption{Cart positions of five cart-pole systems stabilized locally at an update interval of \SI{10}{\ms} and synchronizing their cart positions (middle and right plot) over the multi-hop low-power wireless network at an update interval of 50\ms. \capt{With synchronization enabled, all five carts move in concert and even try to mimic the temporary disturbance of pendulum A (right plot).}}
	\vspace{-1mm}
	\label{fig:synchronization_position_time_plot}
\end{figure*}

\fakepar{Results}
The left plot in \figref{fig:synchronization_position_time_plot} shows the cart positions over time without synchronization.
We see that the carts of the real pendulums move with different amplitude, phase, and frequency due to slight differences in their physics and imperfect measurements.
The simulated pendulums, instead, are perfectly balanced and behave deterministically as they all start in the same initial state.

In the middle plot of \figref{fig:synchronization_position_time_plot}, we can observe the behavior of the pendulums when they synchronize their cart positions over the wireless multi-hop network.
Now, all five carts move in concert.
The movements are not perfectly aligned because, besides the synchronization, each cart also needs to locally stabilize its pole at $\theta = \SI{0}{\degree}$ and $s = \SI{0}{\cm}$.
Since no message is lost during the experiment, the simulated pendulums all receive the same state information and, therefore, show identical behavior.

This effect can also be seen in our third experiment, shown in the right plot of \figref{fig:synchronization_position_time_plot}, where we hold pendulum A for some time at $s = \SI{-20}{\cm}$.
The other pendulums now have two conflicting control goals: stabilization at $s=\SI{0}{\cm}$ and $\theta=\SI{0}{\degree}$, as well as synchronization while one pendulum is fixed at about $s = \SI{-20}{\cm}$.
As a result, they all move towards this position and oscillate between $s = 0$ and $s = \SI{-20}{\cm}$.
Clearly, this experiment demonstrates that the cart-pole systems influence each other, which is enabled by the many-to-all communication over the wireless multi-hop network.



\vspace{-1mm}
\subsection{Impact of Update Interval}
\label{sec:update_interval}

The next experiment takes a look at the impact of different update intervals (and hence end-to-end delays) on control performance.

\fakepar{Setup}
To minimize effects that we cannot control, such as external interference, we use two nodes close to each other: pendulum A (node 1) is stabilized via a remote controller running on node 2.
We test different update intervals in consecutive runs.
Starting with the smallest update interval of 20\ms that the wireless embedded system can support in this scenario, we increase the update interval in steps of 10\ms until stabilization is no longer possible.

%
%
%

\begin{figure}[!tb]
\begin{subfigure}[t]{0.48\linewidth}
    \centering
    \includegraphics[width=1\textwidth]{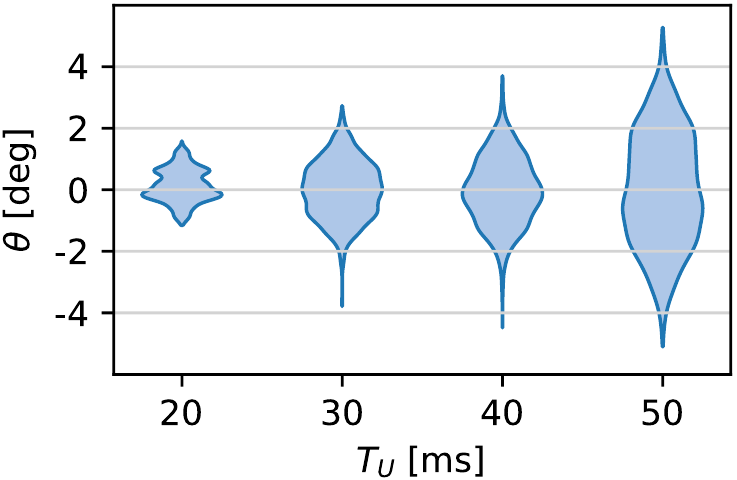}
    \vspace{-4mm}
    \subcaption{Pole angle.}
\end{subfigure}
\,
\begin{subfigure}[t]{0.48\linewidth}
    \centering
    \includegraphics[width=1\textwidth]{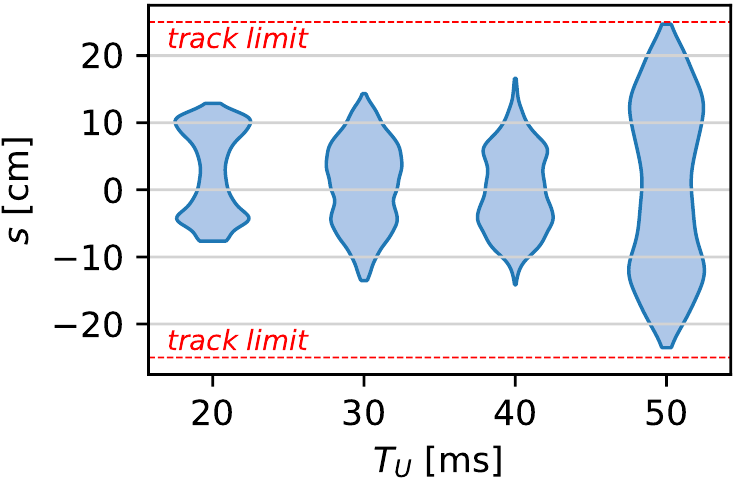}
    \vspace{-4mm}
    \subcaption{Cart position.}
\end{subfigure}
\par\medskip
\begin{subfigure}[t]{0.48\linewidth}
    \centering
    \includegraphics[width=1\textwidth]{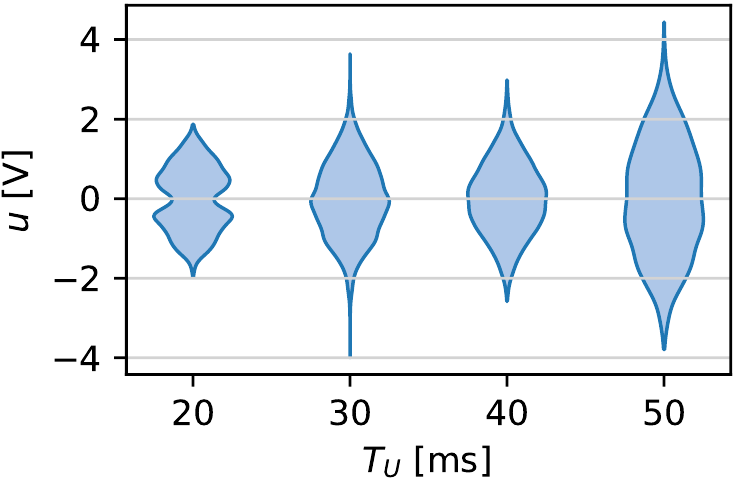}
    \vspace{-4mm}
    \subcaption{Control input.}
\end{subfigure}
\,
\begin{subfigure}[t]{0.48\linewidth}
    \centering
    \includegraphics[width=1\textwidth]{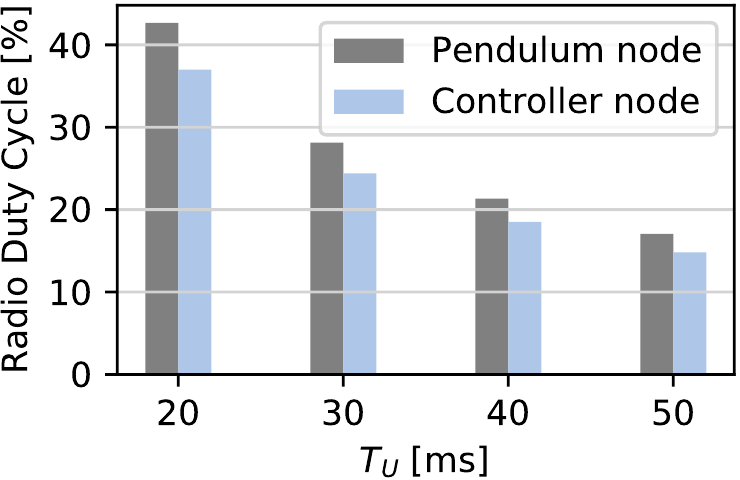}
    \vspace{-4mm}
    \subcaption{Radio duty cycle.}
\end{subfigure}
\vspace{-2mm}
\caption{Distribution of control performance metrics and average radio duty cycle when stabilizing an inverted pendulum over low-power wireless at different update intervals. \capt{A larger update interval leads to larger pole angles and more movement of the cart, but also reduces the average radio duty cycle.}}
\label{fig:timescale}
\vspace{-3mm}
\end{figure}

\fakepar{Results}
\figref{fig:timescale} shows control performance and radio duty cycle for different update intervals based on more than 12,500 data points.
We see that a longer update interval causes larger pole angles and more movement of the cart.
Indeed, the total distance the cart moves during an experiment increases from \SI{3.40}{\m} for \SI{20}{\milli\second} to \SI{9.78}{\m} for \SI{50}{\milli\second}.
This is consistent with the wider distribution of the control input for longer update intervals.
At the same time, the radio duty cycle decreases from 40\percent for \SI{20}{\milli\second} to 15\percent for \SI{50}{\milli\second}.
Hence, there is a trade-off between control performance and energy efficiency, which may be exploited based on the application requirements.

\vspace{-1mm}
\subsection{Resilience to Message Loss}
\label{sec:resilience}

Finally, we evaluate how control performance is affected by message loss, which is a well-known phenomenon in wireless networks~\cite{Srinivasan2010}.

\fakepar{Setup}
We use again the two-node setup from before, but now we fix the update interval at \SI{20}{\ms}.
We let both nodes intentionally drop messages in two different ways.
In a first experiment, the two nodes independently drop a received message according to a Bernoulli process with given failure probability.
Specifically, we test three different failure probabilities in different runs: 15\percent, 45\percent, and 75\percent.
In a second experiment, the two nodes drop a certain number of consecutive messages every \SI{10}{\s}, namely between 10 and 40 messages in different runs.
This artificially violates property \textbf{P2} of the wireless embedded system, yet allows us to evaluate the robustness of our control design to unexpected conditions.



\begin{figure}[!tb]
\begin{subfigure}[t]{0.4\linewidth}
    \centering
    \includegraphics[width=\textwidth]{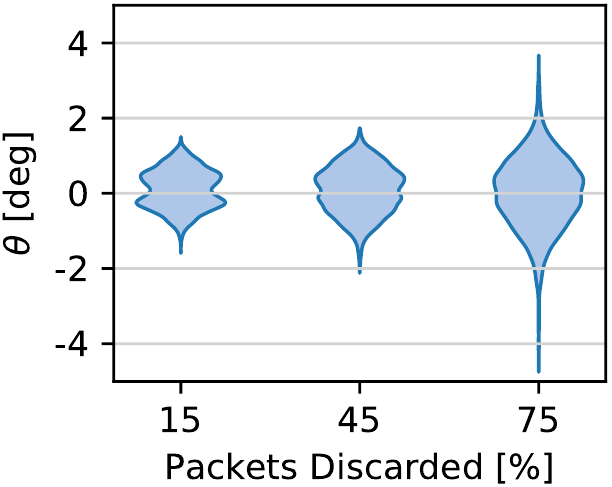}
    \vspace{-4mm}
    \subcaption{Pole angle.}
    \label{fig:drops_angle}
\end{subfigure}
\,
\begin{subfigure}[t]{0.4\linewidth}
    \centering
    \includegraphics[width=\textwidth]{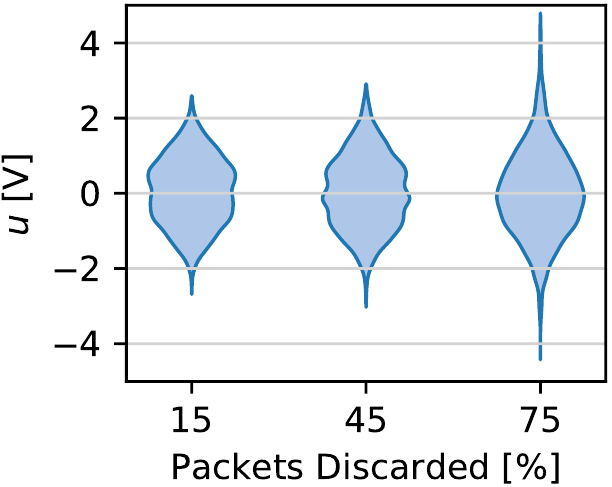}
    \vspace{-4mm}
    \subcaption{Control input.}
    \label{fig:drops_voltage}
\end{subfigure}
\par\medskip
\begin{subfigure}[t]{\linewidth}
    \centering
    \includegraphics[width=\textwidth]{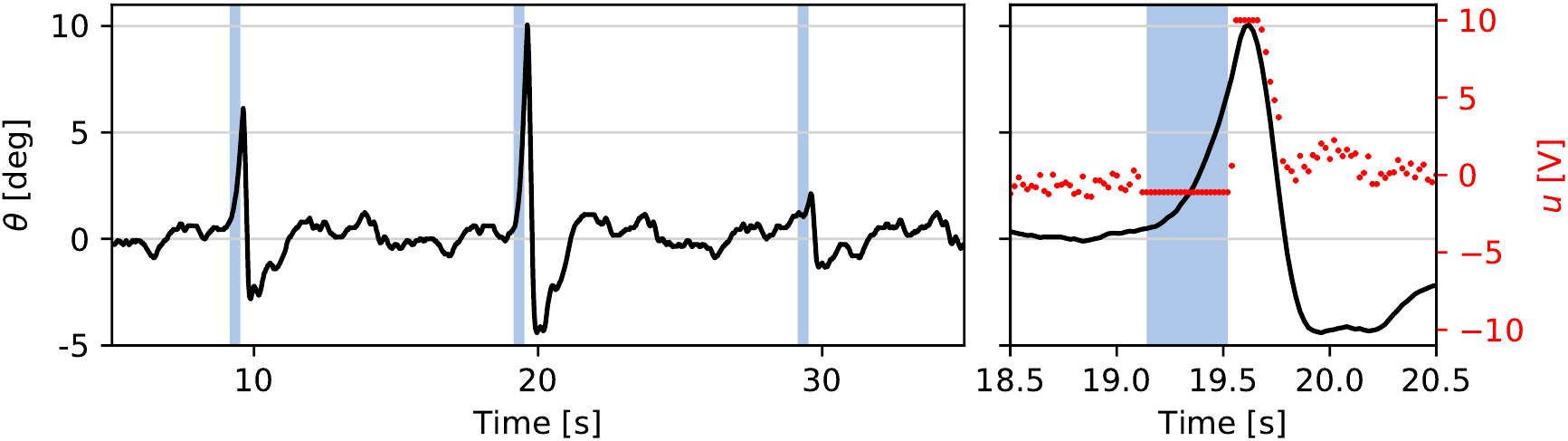}
    \vspace{-4mm}
    \subcaption{Pole angle over time for bursts of 40 consecutive losses every \SI{10}{\s} (shaded areas). The right plot magnifies the second burst.}
    \label{fig:burst}
\end{subfigure}
\vspace{-2mm}
\caption{Control performance and input when stabilizing one pendulum over low-power wireless under artificially injected message loss, for \iid Bernoulli losses (in a and b) and for longer bursts of multiple consecutive losses (in c). \capt{Depending on the update interval, the pendulum can be stabilized despite significant and bursty message loss, albeit with reduced performance.}}
 \vspace{-3mm}
\label{fig:drops}
\end{figure}

\fakepar{Results}
Figures~\ref{fig:drops_angle} and \ref{fig:drops_voltage} show the distributions of the pole angle and the control input for varying \iid Bernoulli message loss rates.
We see that the control performance decreases for higher loss rates, but the pendulum can be stabilized even at a loss rate of 75\%.
One reason for this is the short update interval.
For example, losing 50\percent of the messages at an update interval of 20\ms is comparable to an update interval of 40\ms without any losses, which is enough to stabilize the pendulums as we know from the previous experiment.

\figref{fig:burst} plots the pole angle as a function of time for a burst length of 40 consecutively lost messages, with the right plot zooming into the time around the second burst phase.
No control inputs are received during a burst, and depending on the state of the pendulum and the control input right before a burst, the impact of a burst may be very different as visible in \figref{fig:burst}.
The magnified plot shows that the pole angle diverges from around \SI{0}{\degree} with increasing speed.
As soon as the burst ends, the control input rises to its maximum value of \SI{10}{\V} in order to bring the pendulum back to a non-critical state, which usually takes \SIrange[range-units=single, range-phrase=-]{1}{2}{\s}.
These results show that while property \textbf{P2} of our wireless embedded system design significantly simplifies control design and analysis, the overall system remains stable even if this property is dramatically violated, which is nevertheless very unlikely as demonstrated in prior work~\cite{Zimmerling2013,Karschau2018}.


\vspace{-1mm}
\section{Conclusions}
\label{sec:ending}
We have presented a \cps design that enables, for the first time, fast closed-loop control over multi-hop low-power wireless networks with update intervals of \SIrange[range-units=single, range-phrase=-]{20}{50}{\milli\second}.
Other existing solutions for feedback control over wireless are either limited to the single-hop case or to systems with slow dynamics, where update intervals of several seconds are sufficient.
Through a tight co-design approach, we tame network imperfections
and take the resulting properties of the communication network into account in the control design.
This enables to formally prove closed-loop stability of the entire \cps.
Experiments on a novel cyber-physical testbed with multiple physical systems further demonstrate the applicability, versatility, and robustness of our design.
By demonstrating how to close feedback loops quickly and reliably over multiple wireless hops, this paper is an important stepping stone toward realizing the \cps vision.


\begin{acks}
We thank Harsoveet Singh and Felix Grimminger for their help with the cyber-physical testbed, and the TEC group at ETH Zurich for the design of the DPP platform and making it available to the public. This work was supported in part by the \grantsponsor{dfg}{German Research Foundation}{}~(DFG) within the Cluster of Excellence cfaed (grant \grantnum{dfg}{EXC 1056}), SPP 1914 (grants \grantnum{dfg}{ZI 1635/1-1} and \grantnum{dfg}{TR 1433/1-1}), and the Emmy Noether project NextIoT (grant \grantnum{dfg}{ZI 1635/2-1}), the Cyber Valley Initiative, and the Max Planck Society.
\end{acks}

\balance
\bibliographystyle{ACM-Reference-Format}
\bibliography{references_short}

\vsquish{-2mm}
\appendix
\section{Control Details}
\label{sec:ctrl_details}
In this appendix, we provide further details of the control design and analysis.
We present the proof of Theorem~\ref{thm:MSSourSystem}, implementation details of the controllers we use for the multi-hop stabilization experiments, and outline the approach to multi-agent synchronization.

\vsquish{-2mm}
\subsection{Proof of Theorem \ref{thm:MSSourSystem}}
\label{sec:app_proof}
For clarity, we reintroduce time index $k$ for $\theta$ and $\phi$.
Following a similar approach as in~\cite{Rich2015}, we transform $\theta(k)$ as \mbox{$\theta(k) = \mu_\theta\left(1-\delta_\theta(k)\right)$} with the new binary random variable
$\delta_\theta(k)\in\{1,1-\sfrac{1}{\mu_\theta}\}$ with
$\mathbb{P}[\delta_\theta(k)=1] = 1-\mu_\theta$ and $\mathbb{P}[\delta_\theta(k)=1-\sfrac{1}{\mu_\theta}] = \mu_\theta$; and analogously for $\phi(k)$ and $\delta_\phi(k)$.
We thus have that $\delta_\theta(k)$ is \iid because $\theta$ is \iid with $\E[\delta_\theta(k)] = 0$ and $\Var[\delta_\theta(k)] = \sigma_{p_1}^2$, and similarly for $\delta_\phi(k)$.  Employing this transformation, $\tilde{A}(k)$ in \eqref{eqn:matrix_repr} is rewritten as $\tilde{A}(k) = \tilde{A}_0 + \sum_{i=1}^2 \tilde{A}_i p_i(k)$ with $p_1(k) = \delta_\theta(k)$, $p_2(k)=\delta_\phi(k)$, and $\tilde{A}_i$ as stated in Theorem~\ref{thm:MSSourSystem}.
Thus, all properties of \eqref{eqn:gen_sys} are satisfied, and Lemma \ref{lem:LMIcond} yields the result.

\vsquish{-2mm}
\subsection{Stabilizing Controllers}
\label{sec:app_stab_ctrl}
For the stability experiments of \secref{sec:multihop_stabilization}, we employ the design outlined in \secref{sec:ctrlDesign}.
The system matrices $A$ and $B$ of the cart-pole system that are used for predictions and nominal controller design are given by the manufacturer in~\cite{Quanser2012}.
The nominal controller is designed for an update interval $\Tupdate=\SI{40}{\milli\second}$ via pole placement, and we choose $F$ such that we get closed-loop eigenvalues at \num{0.8}, \num{0.85}, and \num{0.9} (twice).
In experiments with update intervals different from \SI{40}{\milli\second}, we adjust the controller to achieve similar closed-loop behavior.

To derive more accurate estimates of the velocities, filtering can be done at higher update intervals than communication occurs.
For the experiments in \secref{sec:eval}, estimation and filtering occur at intervals between \SI{10}{\milli\second} and \SI{20}{\milli\second}, depending on the experiment.

\vsquish{-2mm}
\subsection{Synchronization}
\label{sec:app_sync}
For simplicity, we consider synchronization of two agents in the following, but the approach directly extends to more than two, as we show in the experiments in \secref{sec:synchronization}.

We consider the architecture in \figref{fig:WirelessControlModel}, where each physical system is associated with a local controller that receives local observations directly, and observations from other agents over the network.  We present an approach based on an optimal LQR~\cite{Anderson2007} to design the synchronizing controllers.
We choose the quadratic cost function
\begin{align}
\label{eqn:cost}
J &= \lim_{K\to\infty}\frac{1}{K}\E\!\Big[\sum\limits_{k=0}^{K-1} \sum_{i=1}^2 \Big(x_i^\mathrm{T}\!(k)Q_i x_i(k) + u_i^\mathrm{T}\!(k)R_i u_i(k) \Big) \nonumber \\
&\phantom{========}+ (x_1(k)-x_2(k))^\mathrm{T}Q_\text{sync} (x_1(k)-x_2(k)) \Big]
\end{align}
which expresses our objective of keeping $x_1(k)-x_2(k)$ small (through the weight $Q_\text{sync}>0$), next to usual penalties on states (\mbox{$Q_i>0$}) and control inputs ($R_i>0$).
Using augmented state $\tilde{x}(k) = (x_1(k), x_2(k))^\mathrm{T}$ and input $\tilde{u}(k) = (u_1(k),$ $u_2(k))^\mathrm{T}$, the term in the summation over $k$ can be rewritten as
\begin{align*}
\tilde{x}^\mathrm{T}\!(k)
\begin{pmatrix}
Q_1+Q_\text{sync}&-Q_\text{sync}\\
-Q_\text{sync}&Q_2+Q_\text{sync}
\end{pmatrix}
\tilde{x}(k)
 +\tilde{u}^\mathrm{T}(k)
 \begin{pmatrix}
R_1&0\\
0&R_2
\end{pmatrix}
 \tilde{u}(k).
\end{align*}
Thus, the problem is in standard LQR form and can be solved with standard tools \cite{Anderson2007}.  The optimal stabilizing controller that minimizes \eqref{eqn:cost} has the structure
$u_1(k) = F_{11} x_1(k) + F_{12} x_2(k)$ and $u_2(k) = F_{21} x_1(k) + F_{22} x_2(k)$;
that is, agent~1 ($u_1(k)$) requires state information from agent~2 ($x_2(k)$), and vice versa.
Because of many-to-all communication, the wireless embedded system directly supports this (as well as any other possible) controller structure (\textbf{P3}).

As the controller now runs on the node that is co-located with the physical process, local measurements and inputs are not sent over the wireless network and the local sampling time 
can be shorter than the update interval of the network over which the states of other agents are received.
While the analysis in \secref{sec:stabAnalysis} can be generalized to the synchronization setting, a formal stability proof is beyond the scope of this paper.
In general, stability is less critical here because of shorter update intervals in the local feedback loop.

For the synchronization experiments in \secref{sec:synchronization}, we choose $Q_i$ in~\eqref{eqn:cost} for all pendulums as suggested by the manufacturer in~\cite{Quanser2012} and set $R_i = 0.1$.
As we here care to synchronize the cart positions, we set the first diagonal entry of $Q_\text{sync}$ to \num{5} and all others to 0.

\end{document}